\documentclass[12pt]{article}
\usepackage{amsmath} 
\usepackage{pdfpages}
\usepackage{graphicx}
\usepackage{float}
\usepackage{authblk}
\usepackage{tabularray}
\usepackage{natbib}

\usepackage{verbatim,caption}
\usepackage{listings}

\usepackage{tikz}
\usetikzlibrary{shapes,arrows}
\usetikzlibrary{arrows,positioning}

\newtheorem{pro}{Proposition}[section]

\newtheorem{defn}{Definition}[section]

\newcommand{\bfa}[1]{\mbox{\boldmath $ #1 $}}
\newcommand{\D}{\displaystyle}

\newfloat{Code}
\begin{document}
 
 \title{Structural identifiability analysis of epidemic models based on differential equations: A tutorial-based primer}

\author[1*]{Gerardo Chowell}
\author[1]{Sushma Dahal}
\author[2]{Yuganthi R. Liyanage}
\author[1]{Amna Tariq}
\author[2]{Necibe Tuncer}

\affil[1]{School of Public Health,  Georgia State University, Atlanta, Georgia, USA}
\affil[2]{Department of Mathematical Sciences, Florida Atlantic University, Boca Raton, Florida, USA

*corresponding author: gchowell@gsu.edu}

\maketitle
  
\begin{abstract}
The successful application of epidemic models hinges on our ability to estimate model parameters from limited observations reliably. An often-overlooked step before estimating model parameters consists of ensuring that the model parameters are structurally identifiable from the observed states of the system. In this tutorial-based primer, intended for a diverse audience, including students training in dynamic systems, we review and provide detailed guidance for conducting structural identifiability analysis of differential equation epidemic models based on a differential algebra approach using DAISY (Differential Algebra for Identifiability of SYstems) and \textit{Mathematica} (Wolfram Research). This approach aims to uncover any existing parameter correlations that preclude their estimation from the observed variables. We demonstrate this approach through examples, including tutorial videos of compartmental epidemic models previously employed to study transmission dynamics and control. We show that the lack of structural identifiability may be remedied by incorporating additional observations from different model states, assuming that the system's initial conditions are known, using prior information to fix some parameters involved in parameter correlations, or modifying the model based on existing parameter correlations. We also underscore how the results of structural identifiability analysis can help enrich compartmental diagrams of differential-equation models by indicating the observed state variables and the results of the structural identifiability analysis.

\end{abstract}

\subsection*{Keywords:} 
Structural identifiability; epidemic models; parameter correlations; DAISY; differential algebra; differential equations.

\subsection*{Acknowledgements}
G.C. is partially supported from NSF grants 2125246 and 2026797 and R01 GM 130900.

\newpage

 \section{Introduction}
 
Mathematical models based on differential equations have become popular tools to investigate infectious disease dynamics and the impact of control interventions during epidemics (see, e.g., 
 \citep{brauer2006some, arino2006simple,chowell2004model,
 chowell2006modelling, anderson2004epidemiology, anderson2020will,gumel2004modelling}). However, their successful application hinges on our ability to reliably estimate model parameters from observations stemming from a limited subset of the epidemiological states of the system (e.g., time series of new cases and deaths) 
 \citep{yan2019quantitative,zhang2021integrated,banks2009mathematical,sauer2021identifiability,gallo2012lack, piazzola2021note}. Estimating parameters for epidemic models is often challenging since models frequently involve several unknown parameters and incorporate epidemiological states that are often not directly observed. In particular, an often-overlooked step before estimating parameters from data consists in conducting structural identifiability analysis to ensure that the model parameters are structurally identifiable \citep{tuncer2018structural,cobelli1980parameter, zhang2021integrated,eisenberg2013identifiability}. This step is crucial since ignoring the structural identifiability of the model parameters can lead to erroneous inferences and public health policies \citep{tuncer2018structural, Gerardo_id,cobelli1980parameter, villaverde2016structural}.\\

Several methods have been proposed to analyze the structural identifiability of model parameters based on differential equations, including the Taylor series method \citep{pohjanpalo1978system}, the generating series method \citep{walter1982global}, the similarity transformation approach \citep{vajda1989similarity}, the direct test method \citep{denis2001some}, and the differential algebra method \citep{ljung1994global,bellu2007daisy}. In particular, methodology based on the differential algebra approach, which is the focus of our tutorial, aims to uncover existing correlations among the model parameters, which may preclude their estimation from the observed variables \citep{tuncer2018structural, TuncerCovid}. When structural identifiability analysis reveals that two or more parameters are correlated, estimating them from the observed variables is impossible because many parameter combinations can produce the same observations, rendering the parameters unidentifiable. As we will illustrate in this tutorial, structural identifiability issues can be resolved in various ways, including reformulating the model to reduce the number of states or parameters or observing additional states of the system. It is important to note that structurally identifiable parameters may still be non-identifiable in practice due to data features \citep{Gerardo_id}. Thus, “practical identifiability”, which is outside the scope of this tutorial, also considers real-world data issues: amount and noise in the data and sampling frequency (e.g., data collection process) \citep{Gerardo_id, zhang2021integrated, villaverde2016structural, TuncerCovid}. Hence, structural non-identifiability implies practical non-identifiability. More importantly, we underscore how structural identifiability analysis can help enrich compartmental diagrams of differential-equation models by providing additional information to the modeler, including the observed variables (e.g., new cases or new deaths) and whether model parameters are structurally identifiable. \\

 We review and provide detailed guidance for conducting structural identifiability analysis of differential equation epidemic models based on the differential algebra approach using DAISY (Differential Algebra for Identifiability of SYstems) \citep{bellu2007daisy} and \textit{Mathematica} (Wolfram Research). It is worth mentioning that DAISY has also been used in structural identifiability of other types of models used in pharmacokinetics \citep{ogungbenro2011structural}, biology \citep{saccomani2010examples}, and engineering \citep{chatzis2015observability}, among others. The differential algebra approach allows the derivation of the model system as a function of the parameters and the observed states of the system, eliminating the unobserved state variables. The power of this method lies in its ability to uncover mathematical expressions describing any existing parameter correlations, which reveal the parameters that are not structurally identifiable from the observed states of the system. As such, structural identifiability analysis yields a theoretical result for the ideal scenario assuming that the observed state variables are provided continuously over time and are noise-free \citep{tuncer2018structural}.\\

In the next section, we start this tutorial by providing an overview of the differential algebra methodology and the input DAISY code. We then supply detailed guidance for their application in a series of epidemic models with varying levels of complexity and assuming that one or more states of the system are observed. The input codes and results from DAISY and \textit{Mathematica} used in this tutorial are publicly available from our GitHub site (https://github.com/sushmadahal/Identifiability). The GitHub repository also contains tutorial videos describing model inputs and results for each model covered in this tutorial. A tutorial video that explains how to install the DAISY software is also available from the GitHub site.

\section{A differential algebra approach to investigate structural parameter identifiability}

To define structural identifiability, we must underscore that observations vary as the model parameters vary. Suppose another set of parameters, denoted by $\hat{\bfa p}$, produced the same observations denoted by $y_1(t;\bfa p)$. That is,  
$$y_1(t;\bfa p) = y_1(t; \hat{\bfa p}).$$
This could only happen when two parameter sets are identical if the model parameters are structurally identifiable \citep{guillaume2019introductory, cobelli1980parameter, bellman1970structural, distefano1980parameter}. That is $\bfa p = \hat{\bfa p}$. Hence, we define the structural identifiability of the model parameters as follows: 
\begin{defn}
Let $\bfa p$ and $ \hat{\bfa p}$ be distinct model parameters, and let $y_1(t;\bfa p)$ be the observations. If
$$ \quad y_1(t;\bfa p) = y_1(t; \hat{\bfa p}) \quad \text{implies}\quad \bfa p = \hat{\bfa p},$$
then we conclude that the model is structurally identifiable from noise-free and continuous observations $y_1(t)$.
\end{defn}

Structural identifiability is an attribute of the model that depends on the structural identifiability of each model parameter. In simple terms, the structural identifiability analysis states that two distinct observations can only be possible with two different sets of parameters. A review of the existing methodologies for structural identifiability analysis is outside the scope of this tutorial. However, we refer the reader to various methods that have been devised to investigate the structural identifiability, including the Taylor series method \citep{pohjanpalo1978system}, the generating series method \citep{walter1982global}, the similarity transformation approach \citep{vajda1989similarity}, the direct test method \citep{denis2001some}, and the differential algebra method \citep{ljung1994global,bellu2007daisy}. Here, we will use the differential algebra approach since it yields additional information regarding existing parameter correlations, which, as we illustrate through various examples, may help find ways to make the model parameters structurally identifiable.\\

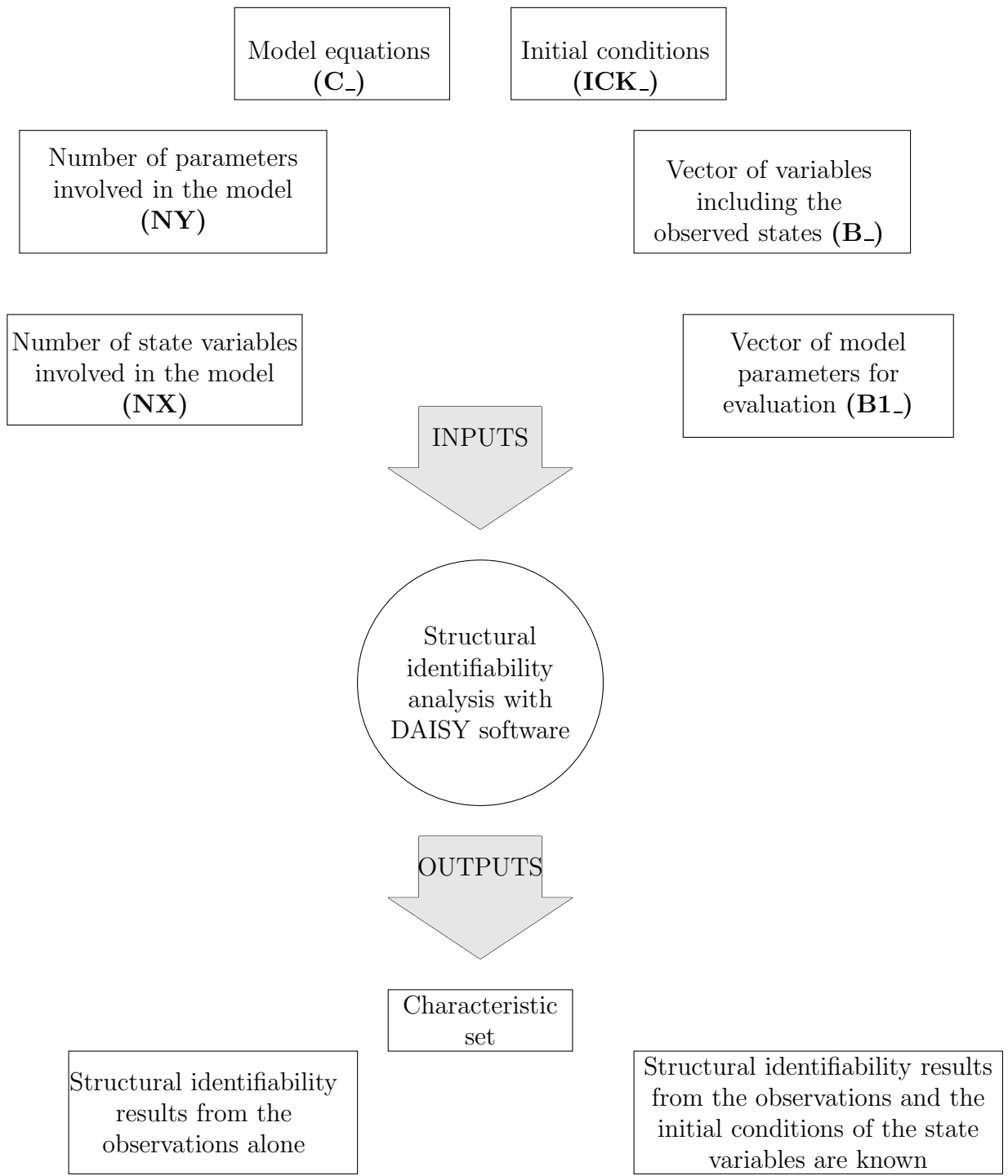
\begin{figure}
    \begin{tikzpicture}
        \draw (-0.2,0) rectangle (4.5,2);
        \node[align=center] at (2,1) {Structural identifiability\\results from the \\observations alone};

        \draw (5,2) rectangle (8,3);
        \node[align=center] at (6.5,2.5) {Characteristic\\set};

         \draw (9,0) rectangle (15,2);
        \node[align=center] at (12,1) {Structural identifiability results\\ from the observations and the\\initial conditions of the state\\variables are known};
         
         \draw[-] (6.5,3.5) --(8,4.5);
          \draw[-] (6.5,3.5) --(5,4.5);
          \draw[-] (5,4.5) --(5.5,4.5);
          \draw[-] (8,4.5) --(7.5,4.5);
          \draw[-] (5.5,4.5) --(5.5,5.5);
          \draw[-] (7.5,4.5) --(7.5,5.5);
           \draw[-] (5.5,5.5) --(7.5,5.5);
           \fill[gray!20] (6.5,3.5) -- (8,4.5) -- (7.5,4.5) -- 
           (7.5,5.5) -- (5.5,5.5) -- (5.5,4.5) -- (5,4.5)--cycle;
           \node[align=center] at (6.5,5) {OUTPUTS};  

            \draw (6.5,8) circle (2);
            \node[align=center] at (6.5,8) {Structural\\ identifiability\\analysis with\\DAISY software}; 

            \draw[-] (6.5,10.5) --(8,11.5);
            \draw[-] (6.5,10.5) --(5,11.5);
            \draw[-] (5,11.5) --(5.5,11.5);
            \draw[-] (8,11.5) --(7.5,11.5);
            \draw[-] (5.5,11.5) --(5.5,12.5);
            \draw[-] (7.5,11.5) --(7.5,12.5);
            \draw[-] (5.5,12.5) --(7.5,12.5);
          \fill[gray!20] (6.5,10.5) -- (8,11.5) -- (7.5,11.5) -- 
          (7.5,12.5) -- (5.5,12.5) -- (5.5,11.5) -- (5,11.5)--cycle;
          \node[align=center] at (6.5,12) {INPUTS};

         \draw (-1.2,12.2) rectangle (3.6,14);
        \node[align=center] at (1.2,13) {Number of state variables\\ involved in the model\\\textbf{(NX)}};

        \draw (-0.3,15) rectangle (4.0,17);
        \node[align=center] at (1.8,16) {Number of observed\\ outputs \textbf{(NY)}};

        \draw (0.7,17.5) rectangle (4.7,19);
        \node[align=center] at (2.6,18) {Model equations\\\textbf{(C\_)}};

         \draw (6.8,17.5) rectangle (12.9,19);
        \node[align=center] at (10.0,18.3) {Vectors of known \textbf{(ICK\_)} and \\unknown \textbf{(ICUNK\_)} \\initial conditions};

        \draw (8.5,15) rectangle (13.5,16.8);
        \node[align=center] at (11,15.8) {Vector of variables\\including the\\ observed states \textbf{(B\_)}};

        \draw (9.5,12) rectangle (14.4,14);
        \node[align=center] at (12,13) {Vector of model\\ parameters for\\evaluation \textbf{(B1\_)}}; 
            
    \end{tikzpicture}
    \caption{Schematic diagram that shows the input and output components associated with structural identifiability analyses conducted using differential algebra with DAISY software.}
    \label{figure_inputOutput}
\end{figure}

Next, we briefly overview the differential algebra method covered in this tutorial. Nevertheless, we refer the reader to review papers detailing the fundamentals of the differential algebra approach to structural identifiability analysis \citep{miao2011identifiability, meshkat2009algorithm}. The differential algebra approach requires several inputs: The model equations, the observed states of the system (observations), the set of model parameters, and whether the initial conditions of the state variables are known. Here, we focus on epidemic models based on systems of nonlinear differential equations. A schematic diagram with the input and output components for structural identifiability analysis with DAISY is shown in Figure \ref{figure_inputOutput}.\\

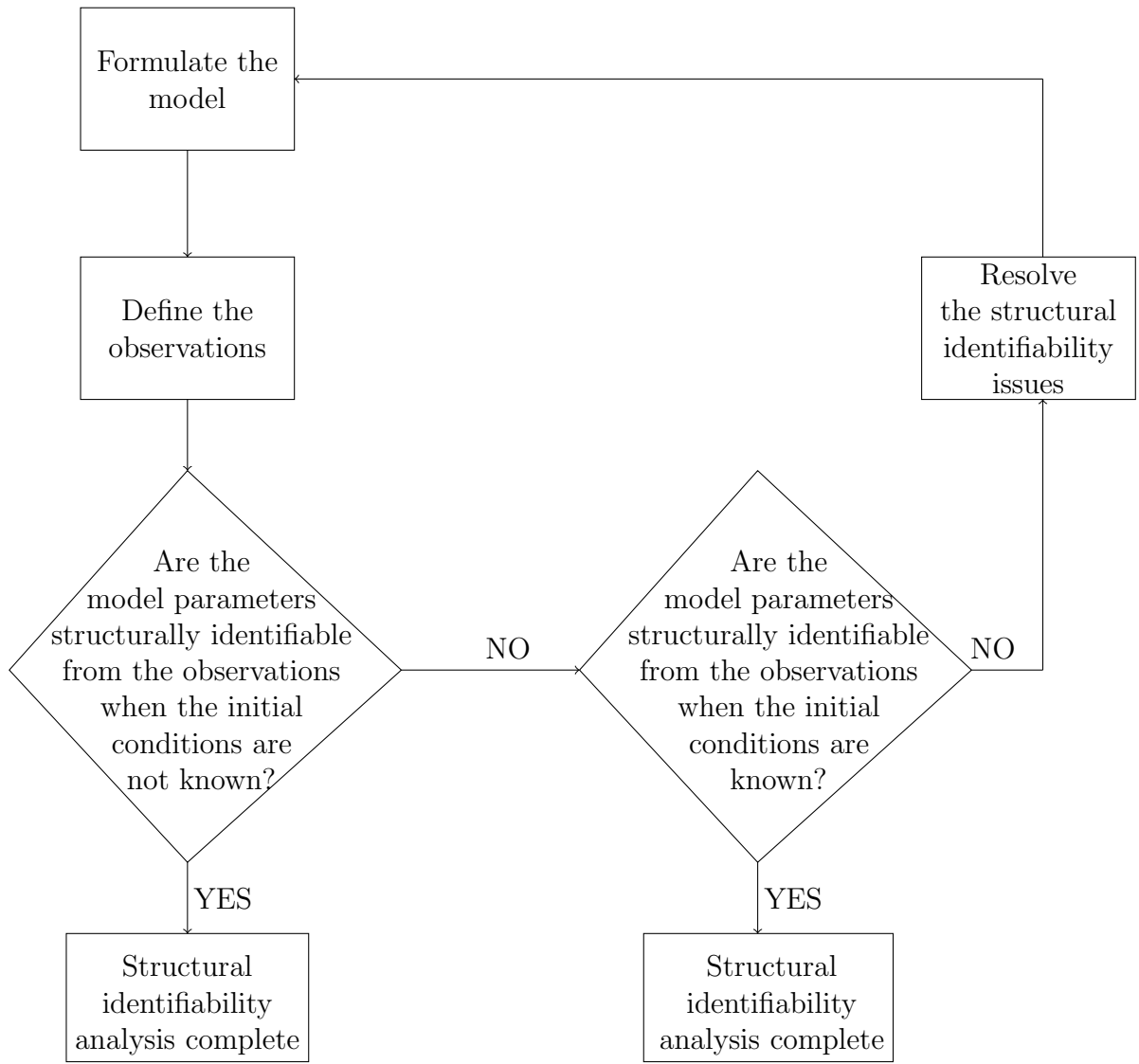
\begin{figure}
    \begin{tikzpicture}
        \draw (-0.2,-0.3) rectangle (3.2,1.5);
        \node[align=center] at (1.5,0.5) {Structural\\identifiability\\analysis complete};
       
       \draw (-1,5.2) -- (1.5,8) -- (4.5,5.2) -- (1.5,2.5) -- cycle;
        \node[align=center] at (1.7,5.2) {Are the\\model parameters\\structurally identifiable\\from the observations\\when the initial\\conditions are \\not known?};
      
       \draw[->] (1.5,2.5) -- (1.5,1.5);
      \node at (2,2) {YES};

       \draw[->] (4.5,5.2) -- (7,5.2);
      \node at (6,5.5) {NO};

        \draw (7,5.2) -- (9.5,8) -- (12.5,5.2) -- (9.5,2.5) -- cycle;
       \node[align=center] at (9.8,5.2) {Are the\\model parameters\\structurally identifiable\\from the observations\\ when the initial\\conditions are\\known?};

       \draw[->] (9.5, 2.5) -- (9.5,1.5);
      \node at (10,2) {YES};

       \draw (7.9,-0.3) rectangle (11.4,1.5);
        \node[align=center] at (9.7,0.5) {Structural\\identifiability\\analysis complete};

        \draw (0,9) rectangle (3,11);
        \node[align=center] at (1.5,10) {Define the\\observations};
        \draw[->] (1.5,9) -- (1.5,8);

      \draw (0,12.5) rectangle (3,14.5);
      \node[align=center] at (1.5,13.5) {Formulate the\\model};
      \draw[->] (1.5,12.5) -- (1.5,11);    

      \draw[-] (12.5,5.2) -- (13.5,5.2);

      \draw[->] (13.5,5.2) -- (13.5,9);

       \node at (12.8,5.5) {NO};
      \draw (11.8,9) rectangle (14.8,11);
        \node[align=center] at (13.3,10) {Resolve \\ the structural\\identifiability\\ issues};
        \draw[-] (13.5,11) -- (13.5,13.5);
        \draw[->] (13.5,13.5) -- (3, 13.5) ;     
    \end{tikzpicture}
    \caption{Schematic diagram of the workflow for conducting structural identifiability analysis.}
    \label{figure_workflow}
\end{figure}

    Using differential algebraic manipulations, the goal is to eliminate unobserved variables and inputs to express the model in terms of observables (outputs) and parameters. An infinite number of differential polynomials and auto-reduced sets can be generated through algebraic manipulation as described in ref. \citep{saccomani2008daisy}. In particular, the set that ranks lowest among these autoreduced sets is the most significant, the characteristic set. The input-output relation consists of differential polynomials (as many as the number of observed outputs) whose variables are given in terms of the observations and their derivatives and whose coefficients are the model parameters. The input-output equations are used for structural identifiability analyses without considering the initial conditions. In contrast, when the initial conditions are known, all of the polynomials of the characteristic set are evaluated at time $0$. Obtaining the characteristics set is a rather complex procedure that usually involves symbolic computation software. To that end, we use \textit{Mathematica} and DAISY software \citep{bellu2007daisy}. It is important to note that the characteristic set of the system of differential equations is not unique and depends on the ranking of the variables. The default ranking in DAISY is that the observations have the lowest ranking, followed by the other state variables in the order specified in the DAISY code. In this case, the first equation in the characteristic set involves only the lowest-ranked variable, the observed variable. The next equation in the characteristic set involves the two lowest-ranked variables. The third equation involves the three lowest-ranked variables, and so on. In summary, the characteristic set serves as a simplified representation of the original model’s differential equations, which allows researchers to determine the structural identifiability of the parameters based on the input-output equations of the model. For a concise introduction to differential algebra, we refer the reader to ref. \citep{bellu2007daisy}.\\

To determine the structural identifiability of the model parameters, DAISY evaluates numerically the characteristic set to assess any existing parameter correlations, assuming whether the model's initial conditions are known or not. Resulting correlation expressions involving two or more parameters will indicate that those parameters are not structurally identifiable, and the modeler will need to devise ways to make the model structurally identifiable. We note that DAISY does not consider that some of the parameters could be functions of the initial conditions. For this reason, the modeler needs to examine the parameter correlations derived by DAISY to fully evaluate the role of the initial conditions on the structural identifiability of the model parameters. As we illustrate in this tutorial with several examples, structural identifiability issues can be resolved by revising the structure of the model, obtaining prior knowledge on specific parameters involved in the parameter correlations, or collecting additional data on particular system states. Figure \ref{figure_workflow} overviews the structural identifiability analysis workflow. The following subsection defines the different user variables involved in the input DAISY code.

\subsection{Overview of the input DAISY code}

Table \ref{Table 0} lists the names of user variables employed in the input DAISY code. Specifically, the user needs to specify the model equations comprising the ordinary differential equation (ODE) system in vector variable \textbf{C\_}. \textbf{Y1} defines the observations from the system. The vector \textbf{B\_} defines the ranking of the variables, including the observations vector \textbf{Y1}. The default ranking in DAISY is that the observations have the lowest ranking, and the other state variables follow the order given in the model definition code. The vector \textbf{B1\_} indicates the unknown parameters we want to identify from the observations \textbf{Y1}. The variables $NX$ and $NY$ indicate the number of state variables and observed outputs in the model, respectively. DAISY can give the structural identifiability results both with and without assumptions of known  initial conditions.\\

We refer the reader to the video recordings (provided on our GitHub site) associated with the examples covered in the tutorial for executing DAISY's input code to conduct structural identifiability analyses.

 \begin{table}[h!]
    \begin{center}
    \footnotesize
    \hspace*{-2cm}
    \begin{tabular}{l|l|l}
    \textbf{User variable}&\textbf{Definition}\\
        \hline
        $\mathbf{B\_}$ & Vector of state variables including the observed states \\ \hline
        $\mathbf{Y1}$ & Vector of observed variables from the model system\\ \hline
        $\mathbf{B1\_}$ & Vector of model parameters for evaluation\\ \hline
       $\mathbf{C\_}$ & Vector with the model equations specifying the ODE system\\ \hline
       $\mathbf{ICK\_}$ & Vector with known initial conditions of the ODE system\\ \hline
       $\mathbf{ICUNK\_}$ & Vector with unknown initial conditions of the ODE system\\ \hline
       ${NX}$ & Number of state variables involved in the ODE system\\ \hline
       ${NY}$ & Number of observed outputs in the system\\ \hline

        \end{tabular}
    \end{center}
    \caption{Variables and their definitions as used in DAISY input code for structural identifiability analysis.}
    \label{Table 0}
\end{table}

\section{Structural identifiability analysis of epidemic models}

We will demonstrate the differential algebra approach to assess the structural identifiability of model parameters through a series of epidemic models with varying levels of complexity. For quick access, Tables \ref{Table 1} and \ref{Table 2} collect the definitions of the state variables and parameters involved in the epidemic models. Summaries of the structural identifiability results for each model when the initial conditions are known and not known are given in Tables \ref{Table 3} and \ref{Table 4}. Tutorial videos as well as input and output DAISY codes for all models analyzed in the tutorial are available from our GitHub site (https://github.com/sushmadahal/Identifiability). \\

We first analyze the simple SEIR (Susceptible-Exposed-Infected-Recovered) model given by the following system of differential equations \citep{brauer2019mathematical}:

\begin{equation}\label{Model1}\tag{M$_1$}
\textbf{Model 1 (M$_1$):}
\begin{cases}
\displaystyle\frac{dS}{dt} = -\beta\frac{SI}{N}, \quad S(0)=S_0\\[1.5ex]
\displaystyle \frac{dE}{dt} = \beta\frac{SI}{N} - kE,\quad E(0)=E_0  \\[1.5ex]
\displaystyle \frac{dI}{dt} = kE-\gamma I, \quad I(0)=I_0 \\[1.5ex]
\displaystyle \frac{dR}{dt} = \gamma I, \quad R(0)=R_0.
\end{cases}
\end{equation}
This model assumes that individuals mix randomly so that each individual has the same probability of contacting any other individual in the population (i.e., homogeneous mixing). The model equations keep track of four epidemiological states over time: the number of susceptible individuals ($S(t)$), latent individuals ($E(t)$), infected individuals ($I(t)$), and recovered individuals ($R(t)$). In model \ref{Model1}, new infections are modeled with a standard incidence term, $\beta\frac{S(t)I(t)}{N}$. A susceptible individual becomes exposed (class $E$) when they get into contact with infectious individuals at the rate $\beta I/N$ where $I(t)/N$ can be interpreted as the probability that a susceptible individual encounters an infectious individual in the population. After the latent period ends ($1/k$), the exposed individual is assumed to be infectious. Infected individuals recover at rate $\gamma$. The model has three epidemiological parameters: the transmission rate ($\beta$), the latent period ($k$), and the recovery rate ($\gamma$). The total population size ($N(t)$) corresponds to a constant parameter since the model describes a closed population ($N = S_0 +E_0 +I_0+ R_0$). Hence, the full parameter set of the SEIR model \eqref{Model1} is given by $\bfa p = \left(N,\beta,k,\gamma\right)$. As described in Section 2, in order to assess the structural identifiability of the model parameters, we need to know the observed states of the system. Next, we need to define the observed variables.
 Observations are functions of the epidemic model's state variables. Here, we consider the case when the observations, $y_1(t)$, correspond to the number of new infected cases. That is in terms of the model variables, this corresponds to:
\begin{equation}\label{model1_data}
y_1(t) = k E(t).
\end{equation}

Thus, we first investigate whether the epidemic model \ref{Model1} is structured to reveal the parameters from the observation $y_1(t)$ alone. \\

\begin{figure}
{\scriptsize
\begin{verbatim}
WRITE "MODEL1-SEIR"$

% B_ IS THE VARIABLE VECTOR

B_:={Y1,X1,X2,X3,X4}$
FOR EACH EL_ IN B_ DO DEPEND EL_,T$

%B1_ IS THE UNKNOWN PARAMETER VECTOR

B1_:={beta,k,gamma1,n}$

%NUMBER OF STATES
NX_:=4$
%NUMBER OF INPUTS 
NU_:=0$
%NUMBER OF OUTPUTS
NY_:=1$

%MODEL  EQUATIONS 

C_:= {df(x1,t) = - beta*x1*x3/n,
      df(x2,t) = beta*x1*x3/n- k*x2,
      df(x3,t) = k*x2 - gamma1*x3,
      df(x4,t) = gamma1*x3, 
            y1 = k*x2}$
FLAG_:=1$
daisy()$

%VALUES OF KNOWN INITIAL CONDITIONS
ICK_:={x1=5,x2=1,x3=0,x4=0}$  
%UNKNOWN INITIAL CONDITIONS
ICUNK_:={}$
%DAISY incorporate these conditional initializations 
%into the structural identifiability analysis 
%by calling subroutine CONDINIZ()
CONDINIZ()$
END$
\end{verbatim}
}
\captionof{Code}{Input DAISY code for model \ref{Model1}, which is available from https://github.com/sushmadahal/Identifiability/tree/main/DAISY. The corresponding compartmental diagram for the SEIR model is given in Figure \ref{fig:model1}. In the code, the model equations comprising the ODE system of the model are defined using the vector \textbf{C\_}. Y1 defines the observations from the system, so the code indicates that we are observing case incidence in this structural identifiability analysis. The vector \textbf{B\_)} defines the ranking of the variables, including the observations vector \textbf{Y1}. The default ranking in DAISY is that the observations have the lowest ranking, and the other state variables follow the order given in the model definition ($S$, $E$, $I$, and $R$). The vector \textbf{B1\_} indicates the model parameters that we want to identify from the observations \textbf{Y1}. The variables \textbf{NX} and \textbf{NY} indicate the number of state variables and observed outputs in the model, respectively. DAISY can give the structural identifiability results with and without assuming that the initial conditions are known. A tutorial video that walks the user through this code and the results obtained from DAISY is given on our GitHub site.}
\label{figdaisy}
\end{figure}

 We run the input Code \ref{figdaisy} using DAISY and obtain the characteristic set of the model \ref{Model1}, which is given in Code \ref{figdaisy_res}. As mentioned before, the characteristic set of the system of differential equations depends on the ranking of the variables. The default ranking in DAISY is that the observations have the lowest ranking, and the other state variables follow the order given in the model definition ($S$,$E$,$I$, and $R$). For instance, the ranking for model \ref{Model1} is 
$$ y_1 < S < E < I < R.$$

The first equation in the characteristic set, which is given by {\texttt{aa\_(1)}} in DAISY (see Code \ref{figdaisy_res}), involves only the lowest ranked variable, which is $y_1(t)$. The next equation in the characteristic set, which is given by {\texttt{aa\_(2)}}in DAISY, involves the 2 lowest ranked variables, which are $y_1(t)$ and $S(t)$. Similarly, the third equation involves the variables $y_1(t), S(t)$, and $E(t)$. Since we are first interested in assessing the structural identifiability of the model parameters from the observed variable, $y_1(t)$,  we focus on the first equation of the characteristic set, which is {\texttt{aa\_(1)}}. As given by DAISY, the first equation is not a monic differential polynomial. So, we divide by the highest degree coefficient, $N^2$, and obtain the following input-output equations of the model \ref{Model1}. The input-output equation \eqref{Model1_input_output}, where input is the model parameters, and output is the observation, is a monic differential polynomial of the observed state variable $y_1(t)$ and the model parameters $\bfa p = \left(N,\beta,k,\gamma\right)$. Working with monic polynomials is important due to the mathematical properties and simplifications they bring. 

\begin{figure}
{\scriptsize
\begin{verbatim}
CHARACTERISTIC SET

aa_(1) := df(y1,t,3)**2*df(y1,t)*y1*n**2 + df(y1,t,3)**2*y1**2*k*n**2 - df(y1,t,3)*df(y1,t,2)**2*y1*n**2 - df(
y1,t,3)*df(y1,t,2)*df(y1,t)**2*n**2 + df(y1,t,3)*df(y1,t,2)*df(y1,t)*y1*n**2*(2*gamma1 - k) + 2*df(y1,t,3)*df(
y1,t,2)*y1**2*k*n**2*(gamma1 + k) - df(y1,t,3)*df(y1,t)**3*n**2*(gamma1 + k) + 4*df(y1,t,3)*df(y1,t)**2*y1**2*
beta*n + df(y1,t,3)*df(y1,t)**2*y1*n**2*(gamma1**2 - 2*k**2) + 8*df(y1,t,3)*df(y1,t)*y1**3*beta*k*n + df(y1,t,
3)*df(y1,t)*y1**2*gamma1*k*n**2*(2*gamma1 + k) + 4*df(y1,t,3)*y1**4*beta*k**2*n + df(y1,t,3)*y1**3*gamma1**2*k
**2*n**2 + df(y1,t,2)**3*df(y1,t)*n**2 - df(y1,t,2)**3*y1*n**2*(gamma1 + k) + 2*df(y1,t,2)**2*df(y1,t)**2*k*n
**2 - 3*df(y1,t,2)**2*df(y1,t)*y1**2*beta*n + df(y1,t,2)**2*df(y1,t)*y1*n**2*(gamma1**2 - gamma1*k - 2*k**2) -
 3*df(y1,t,2)**2*y1**3*beta*k*n + df(y1,t,2)**2*y1**2*k*n**2*(gamma1**2 + 2*gamma1*k + k**2) + df(y1,t,2)*df(
y1,t)**3*n**2*( - 2*gamma1**2 - gamma1*k + 2*k**2) + 2*df(y1,t,2)*df(y1,t)**2*y1**2*beta*n*(3*gamma1 - k) + df
(y1,t,2)*df(y1,t)**2*y1*n**2*(gamma1**3 - gamma1**2*k - 3*gamma1*k**2 - 2*k**3) + 2*df(y1,t,2)*df(y1,t)*y1**3*
beta*k*n*(6*gamma1 + k) + df(y1,t,2)*df(y1,t)*y1**2*gamma1*k*n**2*(2*gamma1**2 + 2*gamma1*k + k**2) + 2*df(y1,
t,2)*y1**4*beta*k**2*n*(3*gamma1 + 2*k) + df(y1,t,2)*y1**3*gamma1**2*k**2*n**2*(gamma1 + k) - df(y1,t)**5*beta
*n + df(y1,t)**4*y1*beta*n*( - 4*gamma1 - 3*k) + df(y1,t)**4*n**2*( - gamma1**3 - 2*gamma1**2*k + k**3) + 4*df
(y1,t)**3*y1**3*beta**2 + df(y1,t)**3*y1**2*beta*n*(gamma1**2 - 6*gamma1*k - 6*k**2) + df(y1,t)**3*y1*gamma1*k
*n**2*( - 2*gamma1**2 - 3*gamma1*k - k**2) + 12*df(y1,t)**2*y1**4*beta**2*k + df(y1,t)**2*y1**3*beta*k*n*(3*
gamma1**2 - 4*k**2) - df(y1,t)**2*y1**2*gamma1**2*k**2*n**2*(gamma1 + k) + 12*df(y1,t)*y1**5*beta**2*k**2 + df
(y1,t)*y1**4*beta*gamma1*k**2*n*(3*gamma1 + 2*k) + 4*y1**6*beta**2*k**3 + y1**5*beta*gamma1**2*k**3*n

aa_(2) :=  - df(y1,t,3)*x1*y1*k*n + df(y1,t,2)*df(y1,t)*x1*k*n - df(y1,t,2)*df(y1,t)*y1*n - df(y1,t,2)*x1*y1*k
*n*(gamma1 + k) - df(y1,t,2)*y1**2*k*n + df(y1,t)**3*n + df(y1,t)**2*x1*k*n*(gamma1 + k) + df(y1,t)**2*y1*n*(
gamma1 + k) - 2*df(y1,t)*x1*y1**2*beta*k + 2*df(y1,t)*y1**2*gamma1*k*n - 2*x1*y1**3*beta*k**2 + y1**3*gamma1*k
**2*n

aa_(3) :=  - x2*k + y1

aa_(4) := df(y1,t,3)*y1*n - df(y1,t,2)*df(y1,t)*n + df(y1,t,2)*x3*y1*beta + df(y1,t,2)*y1*n*(gamma1 + k) - df(
y1,t)**2*x3*beta - df(y1,t)**2*n*(gamma1 + k) - df(y1,t)*x3*y1*beta*gamma1 + 2*df(y1,t)*y1**2*beta - x3*y1**2*
beta*gamma1*k + 2*y1**3*beta*k

aa_(5) :=  - df(x4,t)*df(y1,t,2)*y1*beta + df(x4,t)*df(y1,t)**2*beta + df(x4,t)*df(y1,t)*y1*beta*gamma1 + df(
x4,t)*y1**2*beta*gamma1*k - df(y1,t,3)*y1*gamma1*n + df(y1,t,2)*df(y1,t)*gamma1*n - df(y1,t,2)*y1*gamma1*n*(
gamma1 + k) + df(y1,t)**2*gamma1*n*(gamma1 + k) - 2*df(y1,t)*y1**2*beta*gamma1 - 2*y1**3*beta*gamma1*k

MODEL NOT ALGEBRAICALLY OBSERVABLE

PARAMETER VALUES

b2_ := {beta=2,k=3,gamma1=5,n=7}

MODEL PARAMETER SOLUTION(S)

g_ := {{n=(7*beta)/2,gamma1=5,k=3}}

MODEL NON IDENTIFIABLE

IDENTIFIABILITY WITH ALL INITIAL CONDITIONS (IC_)

ic_ := {x1=5,x2=1,x3=0,x4=0}

MODEL PARAMETER SOLUTIONS

gi_ := {{n=(7*beta)/2,gamma1=5,k=3}}

MODEL NON IDENTIFIABLE
\end{verbatim}
}
\captionof{Code}{Partial DAISY output with results for model \ref{Model1}. When the initial conditions are not known, the parameter correlations indicate that $\beta$ and $N$ are correlated, whereas $\gamma$ and $\kappa$ are well identified. Although DAISY still indicates that $\beta$ and $N$ are correlated when the initial conditions are known, we know that $N$ is known if the initial conditions of the system are known since $N=S_0+E_0+I_0+R_0$.}
\label{figdaisy_res}
\end{figure}

\begin{equation}\label{Model1_input_output}
\begin{aligned}
&(y_1^{'''})^2 y_1^{'} y_1 +  (y_1^{'''})^2 y_1^2  k - 
 y_1^{'''} (y_1^{''})^2 y_1  - 
 y_1^{'''} y_1^{''} (y_1^{'})^2  + \\[2.5ex]
& y_1^{'''} y_1^{''} y_1^{'} y_1  (2 \gamma - k) + 
 2 y_1^{'''} y_1^{''} y_1^2 k  (\gamma + k) - 
y_1^{'''} (y_1^{'})^3  (\gamma + k) + \\[2.5ex]
&4 y_1^{'''} (y_1^{'})^2 y_1^2 \frac{\beta}{N} + y_1^{'''} (y_1^{'})^2 y_1  (\gamma^2 - 2 k^2) + 
8 y_1^{'''} y_1^{'} y_1^3 \frac{\beta k }{N} + y_1^{'''} y_1^{'} y_1^2 \gamma k (2 \gamma + k) + \\[2.5ex]
&4 y_1^{'''} y_1^4 \frac{\beta k^2}{ N} + y_1^{'''} y_1^3 \gamma^2 k^2  + 
(y_1^{''})^3 y_1^{'}  - (y_1^{''})^3 y_1  (\gamma + k) + 2 (y_1^{''})^2 (y_1^{'})^2 k  - \\[2.5ex]
& 3 (y_1^{''})^2 y_1^{'} y_1^2 \frac{\beta}{N} +
(y_1^{''})^2 y_1^{'} y_1  (\gamma^2 - \gamma k - 2 k^2) - 
 3 (y_1^{''})^2 y_1^3 \frac{\beta k}{ N }+ (y_1^{''})^2 y_1^2 k  (\gamma^2 + 2 \gamma k + k^2) +\\[2.5ex]
&y_1^{''} (y_1^{'})^3  ( - 2 \gamma^2 - \gamma k + 2 k^2) +  2 y_1^{''} (y_1^{'})^2 y_1^2 \frac{\beta  (3 \gamma - k)}{N} +
y_1^{''}(y_1^{'})^2 y_1  (\gamma^3 - \gamma^2 k - 3 \gamma k^2 - 2 k^3) + \\[2.5ex]
&2 y_1^{''} y_1^{'} y_1^3 \frac{\beta k}{N} (6 \gamma + k) + 
y_1^{''} y_1^{'} y_1^2 \gamma k  (2 \gamma^2 + 2 \gamma k + k^2) + 
2 y_1^{''}y_1^4 \frac{\beta k^2}{N} (3 \gamma + 2 k) + \\[2.5ex]
& y_1^{''} y_1^3 \gamma^2 k^2  (\gamma + k) -  (y_1^{'})^5 \frac{\beta}{N} + 
(y_1^{'})^4 y_1 \frac{\beta  ( - 4 \gamma - 3 k)}{N} + (y_1^{'})^4  ( - \gamma^3 - 2 \gamma^2 k + k^3) + \\[2.5ex]
 & 4 (y_1^{'})^3 y_1^3 \frac{\beta^2}{N^2} + (y_1^{'})^3 y_1^2 \frac{\beta  (\gamma^2 - 6 \gamma k - 6 k^2)}{N} + 
 (y_1^{'})^3 y_1 \gamma k  ( - 2 \gamma^2 - 3 \gamma k - k^2) + \\[2.5ex]
&12 (y_1^{'})^2 y_1^4 \frac{\beta^2 k}{N^2} + (y_1^{'})^2 y_1^3 \frac{\beta k  (3 \gamma^2 - 4 k^2)}{N} - 
(y_1^{'})^2 y_1^2 \gamma^2 k^2  (\gamma + k) + 12 y_1^{'} y_1^5 \frac{\beta^2 k^2}{N^2} + \\[2.5ex]
&y_1^{'} y_1^4 \frac{\beta \gamma k^2(3 \gamma + 2 k)}{N} + 4 y_1^6 \frac{\beta^2 k^3}{N^2} + y_1^5 \frac{\beta \gamma^2 k^3}{ N}=0
\end{aligned}
\end{equation}

It is worth mentioning that for a given set of parameters, $\bfa p = \left(N,\beta,k,\gamma\right)$, solving the monic differential polynomial \eqref{Model1_input_output} is equivalent to solving the model \ref{Model1} in terms of the observed states of the system and the model parameters. For studying structural identifiability, we analyze whether another set of parameters, $\hat{\bfa p} =  \left(\hat N,\hat \beta, \hat k,\hat \gamma\right)$ could produce the same output \eqref{Model1_input_output}. That is, within differential algebra approach, we define structural identifiability as follows:
\begin{defn}\label{defn1}
Let $c(\bfa p)$ denote the coefficients of the input-output equation \eqref{Model1_input_output}  with $\bfa p =  \left(N,\beta,k,\gamma\right)$ denoting the parameters of the model \ref{Model1}. Then we say that the model \ref{Model1} is structured to reveal its parameters from the observations  $y_1(t) = k E(t)$ if and only if 
$$ c(\bfa p) = c(\hat{\bfa p}) \implies \bfa p = \hat{\bfa p}$$
where $\hat{\bfa p} =  \left(\hat N,\hat \beta, \hat k,\hat \gamma\right)$ is another set of parameters assumed to produce the same observations. 
\end{defn}
Here, we list some of the  coefficients of the input-output equation \eqref{Model1_input_output} of model \ref{Model1} as, 
\begin{equation*}
\begin{aligned}
c(\bfa p) = &\left\{ k, \; 2\gamma -k,\; k(\gamma+k),\; \D\frac{\beta}{N},\;(\gamma^2 - 2 k^2),\; k (2\gamma +k),\; \D\frac{\beta k^2}{N},\;  (\gamma^2 - \gamma k - 2 k^2),\; \right.\\
&\left. (2 \gamma^2 + 2 \gamma k + k^2), \; ( - 2 \gamma^2 - \gamma k + 2 k^2),\; \D\frac{\beta  (3 \gamma - k)}{N},\; (\gamma^3 - \gamma^2 k - 3 \gamma k^2 - 2 k^3),\; \right.\\
&\left.  \frac{\beta k}{N} (6 \gamma + k) \right\} 
\end{aligned}
\end{equation*}

DAISY numerically solves $c(\bfa p) = c(\hat{\bfa p})$ by assigning random integers to the parameter values although epidemic model parameters are typically not integers. Examining Code \ref{figdaisy_res}, we see that DAISY assigned the following values for the model parameters
$$ \hat{\bfa p} = \left\{ \hat{\beta} =2,\; \hat{k}=3,\; \hat{\gamma} =5,\; \hat{N}=7\right\}. $$
Then DAISY solves the system $c(\bfa p) = c(\hat{\bfa p})$ for these integers, and gives the following result
\begin{equation}\label{d_res_m1}
    \left\{ N = \D\frac{7\beta} {2},\; k=3,\; \gamma =5,\right\}. 
\end{equation}  
Since parameters $\beta$ and $N$ are correlated, DAISY declares that the model \ref{Model1} is not identifiable (see Code \ref{figdaisy_res}). To consider arbitrary values of the parameters (not just integers), we also solve $c(\bfa p) = c(\hat{\bfa p})$ for arbitrary parameter values using \textit{Mathematica} and obtain the following parameter correlations:
\begin{equation}\label{m_res_m1}
    k =\hat k,\; \gamma = \hat \gamma,\; \D\frac{\beta}{N} = \D\frac{\hat \beta}{\hat N}.
\end{equation} 
which agree with those derived using DAISY \eqref{d_res_m1}. The corresponding \textit{Mathematica} code is available from our GitHub site. This means we can only identify the rate at which individuals progress the latent period, $k$, and the recovery rate, $\gamma$. Still, we cannot identify the transmission rate, $\beta$, or the total population size, $N$. The correlation expression indicates that it is only possible to identify the ratio $\D\frac{\beta}{N}$. That is, any $\beta$ or $N$ values that produce the same ratio $\D\frac{\beta}{N}$, which is determined by the observation, will yield the same observations $y_1(t)$. We summarize the structural identifiability analysis of the model \ref{Model1} from observations $y_1(t)$ in Proposition \ref{prop1}. 

\begin{pro}
\label{prop1}
The epidemic model \ref{Model1} is not structured to identify all of its parameters from the number of new cases given by $y_1(t) = k E(t)$. We can only identify 
$$k, \gamma \text{ and } \D\frac{\beta}{N} $$
\end{pro}
Since, model \ref{Model1} is not structurally identifiable, next step is to resolve this issue using the correlation information obtained through differential algebra approach. Clearly,  model \ref{Model1} can become structurally identifiable if the total population size $N$ is known. Hence, we state the following result.

\begin{pro}
If the total population size $N$ is known, then the model \ref{Model1} is structurally identifiable from the observations of $y_1(t) = k E(t)$. We can identify the remaining three parameters,
$k, \gamma, \beta.$
\end{pro}

\begin{figure}
\caption{Compartmental diagram of model \ref{Model1}. Circles show the epidemiological compartments for the different states of the system. Solid arrows indicate the transitions between compartments. The dashed arrows indicate the source of the observations, which are the newly infected individuals. Results from the structural identifiability analysis are summarized for the scenarios when the initial conditions are known and unknown.}\label{fig:model1}
\tikzstyle{line} = [draw, -latex']
\begin{tikzpicture} 
  \node [circle,draw,minimum size=1cm] (S) at (0,0) {$S$}; 
  \node [circle,draw,minimum size=1cm] (E) at (3,0) {$E$}; 
  \node [circle,draw,minimum size=1cm]  (I) at (6,0) {$I$}; 
  \node [circle,draw,minimum size=1cm]  (R) at (9,0) {$R$};
  \node [draw,rounded rectangle,text width=3cm] (O) at (4.5,-2) {Observations are new cases; $y_1(t) = kE(t)$};
  \path [line]    (S) -- node [midway,above ] {$ \D \frac{\beta S I}{N}$}  (E); 
  \draw [line]   (E) -- node [midway,above,sloped ] (c){$k E $\\[.2em]} (I); 
  \draw [->]     (I) -- node [midway,above,sloped ] {$\gamma I $\\[.2em]}  (R); 
  \draw[->, dashed] (c) --(O);
  \node [draw,text width=6cm] (idr) at (1,-5) {\textbf{Identifiability Results:}\\\textbf{Without Initial Conditions}\\ identifiable parameters: $k,\gamma$ \\ unidentifiable parameters: $\beta,N$\\
  parameter correlations $\D\frac{\beta}{N} =\D\frac{\hat\beta}{\hat N} $};
   \node [draw,text width=6cm] (idr) at (9,-5) {\textbf{Identifiability Results:}\\\textbf{With Initial Conditions}\\ identifiable parameters: $k,\gamma,\beta,N$ \\ };
\end{tikzpicture}
\end{figure}
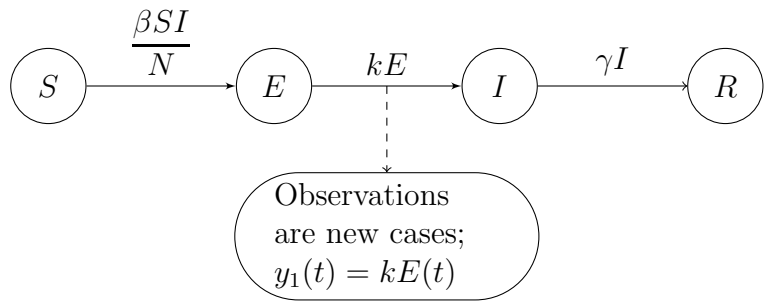

Note that $N = S_0 + E_0 +I_0 +R_0$. This means that total population size $N$ is known if all the initial conditions are given. The DAISY result when the initial conditions are known (see Code \ref{figdaisy_res}) states that the model is not identifiable since DAISY does not take into account the fact that N is a constant in this model and can be derived from the initial conditions. We emphasize the need to carefully examine the parameter correlations derived by DAISY in order to fully evaluate the role of the initial conditions on the structural identifiability of the model parameters. Hence, we state the following proposition.

\begin{pro}
If all the initial conditions are given for the model \ref{Model1}, then the model is structurally identifiable from the observations of $y_1(t) = k E(t)$. We can identify all model parameters,
$k, \gamma, \beta, N.$
\end{pro}

The enhanced compartmental diagram of model \ref{Model1} that indicates the observed states of the system and the parameter correlations resulting from the structural identifiability analysis is given in Figure \ref{fig:model1}.


\subsection{SEIR model with symptomatic and asymptomatic infections}

Next, we consider an extension of the SEIR model that incorporates symptomatic and asymptomatic infections \citep{chowell2006transmission}, which is given by the following system of differential equations:

\begin{equation}\label{Model2}\tag{M$_2$}
	\textbf{Model 2 (M$_2$):}
	\begin{cases}
		\displaystyle \frac{dS}{dt} = -\beta\frac{SI}{N}, \quad S(0)=S_0, \\[1.5ex]
		\displaystyle \frac{dE}{dt} = \beta\frac{SI}{N}-kE, \quad E(0)=E_0,  \\[1.5ex]
		\displaystyle \frac{dI}{dt} = k \rho E-\gamma I, \quad I(0)=I_0, \\[1.5ex]
		\displaystyle \frac{dA}{dt} = k (1-\rho) E-\gamma A, \quad A(0)=A_0, \\[1.5ex]
		\displaystyle \frac{dR}{dt} = \gamma I +  \gamma A, \quad R(0)=R_0.
	\end{cases}
\end{equation}

In this second model, there are five epidemiological states: the number of susceptible individuals ($S(t)$), the number of latent individuals ($E(t)$), the number of symptomatic infectious individuals ($I(t)$), the number of asymptomatic (not infectious) individuals ($A(t)$), and the number of recovered individuals ($R(t)$). Compared to model \ref{Model1}, in this model, we have one additional epidemiological state, $A(t)$. The susceptible individuals upon contact with infected individuals $I(t)$ move to the exposed class at rate $\beta I(t)/N$. The fraction of exposed individuals becomes symptomatic infectious cases at rate $ k \rho$, whereas the remaining exposed individuals move to the asymptomatic class at rate 
$k(1-\rho)$. Hence the proportion $\rho$ is always less than $1$. The infected individuals recover at rate $\gamma$.\\

Since the model describes a closed population, the total population remains constant throughout the epidemic ($N = S_0 +E_0 +I_0+ A_0 +R_0$). The full parameter set of the model \ref{Model2} is given by $\bfa p = \left(N,\beta,k,\rho,\gamma\right)$. Next, we introduce  the observed variables for this example.
We consider the number of newly symptomatic cases as the observations $y_1(t)$. Hence,  for model \ref{Model2}, we define the observations as, 
\begin{equation}\label{model1_data}
y_1(t) = k\rho E(t).
\end{equation}
As before, we use DAISY to obtain the characteristic set and first study the structural identifiability of the model \ref{Model2} when the initial conditions are unknown as outlined in Section 2. The input-output equations \eqref{Model2_input_output} are given in the Appendix and the characteristic set is available from our GitHub site. A close examination of the input-output equation \eqref{Model2_input_output}, reveals that the $\rho$ parameter modeling the proportion of symptomatic infections is absent from the equation. This indicates that the parameter $\rho$ is unidentifiable. We use the Definition \ref{defn1} to determine the identifiability of the remaining parameters. 

Let $c(\bfa p)$ denote the coefficients of the input-output equation \eqref{Model2_input_output}  with $\bfa p =  \left(N,\beta,k,\gamma,\rho\right)$ denoting the parameters of the model \ref{Model2}. Let $\hat{\bfa p} =  \left(\hat N,\hat \beta, \hat k,\hat \gamma, \hat \rho\right)$ be another set of parameters assumed to produce the same observations. Solving $c(\bfa p) = c(\hat{\bfa p})$ (See \textit{Mathematica} file) we obtain the following solution,
$$ k =\hat k,\; \gamma = \hat \gamma,\; \D\frac{\beta}{N} = \D\frac{\hat \beta}{\hat N}.$$ 
The solution indicates that only the parameters $k$ and $\gamma$ are structurally identifiable. Parameter $\rho$ cannot be identified since it does not appear in the input-output equation \eqref{Model2_input_output}. As in model \ref{Model1}, the transmission rate $\beta$ and total population size $N$ are correlated. We summarize the structural identifiability analysis of the model \ref{Model2} from observations of newly symptomatic cases in the Proposition \ref{m2prop1}. 
\begin{pro}\label{m2prop1}
The model \ref{Model2} is not structured to reveal parameters $N$,$\beta$, and $\rho$ from the observations of $y_1(t) = k\rho E(t)$. Only the parameters $k$ and $\gamma$ can be identified. Hence, model  \ref{Model2} is not structurally identifiable.
\end{pro}

When all initial conditions are known, DAISY indicates that parameter $\rho$ can now be identified (see DAISY's output in our GitHub site). We note that $\rho$ does not appear in the input-output equation, which is the first equation of the characteristic set. However, it appears in the other equations of the characteristic set. Moreover, DAISY still finds the correlation between $\beta$ and $N$. Nevertheless, as in the previous example, the total population size $N$ is known from knowledge of the initial conditions since $N = S_0 + E_0 +I_0 +R_0$. Therefore, we summarize the identifiability analysis when the initial conditions are known in Proposition \ref{m2prob2}. 
\begin{pro} \label{m2prob2}
The model \ref{Model2} is structured to identify all its parameters, $\bfa p = \left(N,\beta,k,\gamma,\rho\right)$, from the newly symptomatic cases $y_1(t) = k\rho E(t)$ if all the initial conditions are known.
\end{pro}

\begin{figure}
\caption{Compartmental diagram for model \ref{Model2}. Circles show the epidemiological compartments for the different states of the system. Solid arrows indicate the transitions between compartments. The dashed arrows indicate the source of the observations, which are the newly infected individuals. The results of the structural identifiability analysis are summarized when initial conditions are known and unknown.}\label{fig:model2}
\tikzstyle{line} = [draw, -latex']

\begin{tikzpicture} 
  \node [circle,draw,minimum size=1cm] (S) at (0,0) {$S$}; 
  \node [circle,draw,minimum size=1cm] (E) at (3,0) {$E$}; 
  \node [circle,draw,minimum size=1cm]  (I) at (6,-2) {$I$}; 
  \node [circle,draw,minimum size=1cm]  (A) at (6,2) {$A$}; 
  \node [circle,draw,minimum size=1cm]  (R) at (9,0) {$R$};
  \node [draw,rounded rectangle,text width=3cm] (O) at (4.5,-3.5) {Observations are new cases; $y_1(t) = k \rho E(t)$};
  \path [line]    (S) -- node [midway,above ] {$ \D \frac{\beta S I}{N}$}  (E); 
  \draw [line]   (E) -- node [midway,above,sloped ] (c){$k \rho E $\\[.2em]} (I); 
  \draw [line]   (E) -- node [midway,above,sloped ] {$k (1-\rho) E $\\[.2em]} (A); 
  \draw [->]     (I) -- node [midway,above,sloped ] {$\gamma I $\\[.2em]}  (R); 
  \draw [->]     (A) -- node [midway,above,sloped ] {$\gamma A $\\[.2em]}  (R); 
  \draw[->, dashed] (c) --(O);
  \node [draw,text width=6cm] (idr) at (1,-7) {\textbf{Identifiability Results:}\\\textbf{Without Initial Conditions}\\ identifiable parameters: $k,\gamma$ \\ unidentifiable parameters: $\beta,N, \rho$\\
  parameter correlations $\D\frac{\beta}{N} =\D\frac{\hat\beta}{\hat N} $};
   \node [draw,text width=6cm] (idr) at (9,-7) {\textbf{Identifiability Results:}\\\textbf{With Initial Conditions}\\ identifiable parameters: $k,\gamma,\beta,N, \rho$ \\ };
\end{tikzpicture}
\end{figure}
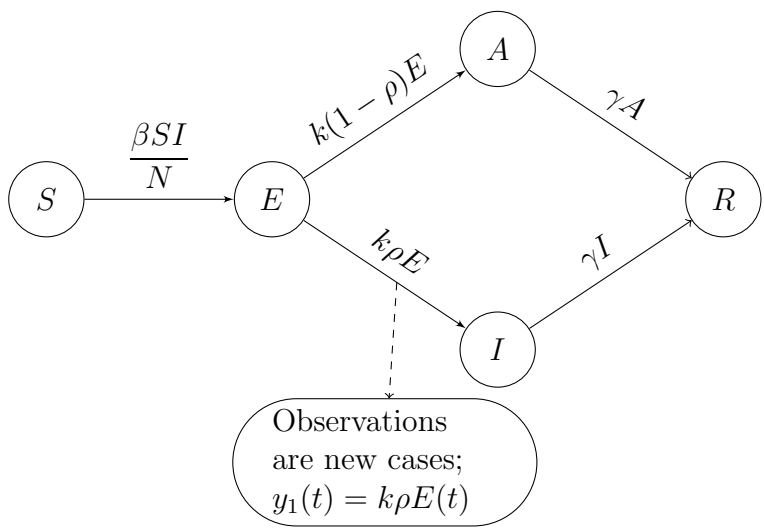

The enhanced compartmental diagram of the model \ref{Model2} includes the observed states of the system and the parameter correlations resulting from the structural identifiability analysis (Figure \ref{fig:model2}).

\subsection{SEIR model with infectious asymptomatic individuals (\ref{Model3})}
Here we consider an extension of the previous model by assuming that asymptomatic individuals are infectious and contribute to the force of infection \citep{chowell2007comparative}. Thus, in model \ref{Model3}, the incidence is modeled as 
$$ \D\frac{(\beta_A A+\beta_I I)S}{N}.$$

where $\beta_A$ is the transmission rate of asymptomatic individuals and $\beta_I$ is the transmission rate of symptomatic individuals. The model is given by the following system of differential equations:

\begin{equation}\label{Model3}\tag{M$_3$}
\textbf{Model 3 (M$_3$):}
\begin{cases}
\displaystyle\frac{dS}{dt} = -\frac{(\beta_A A+\beta_I I)S}{N},  \quad S(0) = S_0 \\[1.5ex]
\displaystyle \frac{dE}{dt} = \frac{(\beta_A A+\beta_I I)S}{N} - kE,  \quad E(0) = E_0\\[1.5ex]
\displaystyle \frac{dI}{dt} = k\rho E-\gamma I, \quad I(0) = I_0 \\[1.5ex]
\displaystyle \frac{dA}{dt} =k(1-\rho) E -\gamma A, \quad A(0)=A_0 \\[1.5ex]
\displaystyle \frac{dR}{dt} = \gamma I+\gamma A, \quad R(0) = R_0. 
\end{cases}
\end{equation}

Compared to model \ref{Model2}, here we distinguish between the transmission rates for symptomatic ($\beta_I$) and asymptomatic infectious ($\beta_A$) individuals. Similar to epidemic model \ref{Model2}, in this model, a fraction of exposed individuals move to the symptomatic class at the rate $ k \rho$, and the remaining fraction move to the asymptomatic class at the rate $k (1-\rho)$. Both symptomatically and asymptomatically infected individuals recover at rate $\gamma$. The full parameter set of the model \ref{Model3} is then $\bfa p = \left(N,\beta_A,\beta_I,k,\rho,\gamma\right)$. 
For this model, we take the  observations to be the number of new symptomatic infections. In terms of the model variables, we define the observation, $y_1(t)$ as;
\begin{equation}\label{model1_data}
y_1(t) = k\rho E(t).
\end{equation}

Using DAISY, we obtain the input-output equation \eqref{Model3_input_output} of the model \ref{Model3}, which is given in the Appendix. The corresponding files with the input DAISY code and results are available from our GitHub site.\\

To determine the identifiability of the model parameters, we use Definition \ref{defn1}. Therefore, set $c(\bfa p) = c(\hat{\bfa p})$, where $\hat{\bfa p} =  \left(\hat N, \hat k,\hat \gamma, \hat \rho, \hat{\beta}_A, \hat{\beta}_I \right)$ is the set of other parameters presumed to yield identical observations as $\bfa p$. We then use \textit{Mathematica} to solve the nonlinear system $c(\bfa p) = c(\hat{\bfa p})$ for $\bfa p$ and obtain the following solution:

 \begin{equation}
    \label{m3res}
 k =\hat k,\; \gamma = \hat \gamma,\; \beta_I \rho +\beta_A (1- \rho) = \hat\beta_I \hat\rho + \hat \beta_A (1-  \hat\rho),\; N \rho =\hat N \hat\rho.
\end{equation}

The solution set \eqref{m3res} indicates that we can only identify parameters $k$ and $\gamma$. Parameters $\rho$, $\beta_A$, $\beta_I$ and $N$ cannot be identified. Indeed, we found two parameter correlations, one involving the transmission rates ($\beta_A$, $\beta_I$) and the fraction of symptomatic infections ($\rho$) and the second one involving the fraction of symptomatic infections ($\rho$) and the total population size ($N$). Thus, we state the following proposition regarding the structural identifiability analysis of model \ref{Model3} from the observations.
\begin{pro}
The epidemic model \ref{Model3} is not structured to reveal parameters $N$,$\rho$,$\beta_A$,$\beta_I$ from the observations of $y_1(t) = k \rho E(t)$. Only parameters, $\gamma$ and $k$ can be identified. Hence, model  \ref{Model3} is not structurally identifiable from the observations of new symptomatic infections. 
\end{pro}

We now compare the \textit{Mathematica} result \eqref{m3res} with DAISY result, which is shown in Code \ref{fig:model3_daisy}. DAISY assigns integers for the parameters. As seen in Code \ref{fig:model3_daisy}, DAISY assigns  $\hat{\bfa p} = \left\{\hat{N} =13,\; \hat{k}=5,\;\hat{\gamma} = 11,\; \hat{\rho}=7,\; \hat{\beta}_A=2,\; \hat{\beta}_I =3 \right\} $.  It is worth mentioning one of the shortcomings of DAISY, which is assigning integers for the parameters.  Clearly, assigning an integer value greater than one for $\rho$ is not realistic (as DAISY assigned $\hat{\rho} =7$). Since $\rho$ is the fraction of infected individuals moving to the symptomatic class, it is clearly less than one. The \textit{Mathematica} result contains two parameter correlations, while DAISY has given the ratio of these two correlations:  
\begin{equation}\label{same_corr}
    \D\frac{ \beta_I \rho +\beta_A (1- \rho)}{N \rho} =  \D\frac{\hat\beta_I \hat\rho + \hat \beta_A (1-  \hat\rho)}{\hat N \hat\rho}
\end{equation} 

\begin{figure}
    \centering
   {\scriptsize
\begin{verbatim}
PARAMETER VALUES$

b2_ := {betaa=2,betai=3,k=5,rho1=7,gamma1=11,n=13}$

MODEL PARAMETER SOLUTION(S)$

g_ := {{n=( - 91*betaa*rho1 + 91*betaa + 91*betai*rho1)/(9*rho1),gamma1=11,k=5}}$



\end{verbatim}
    }
    \captionof{Code}{Parameter correlations derived by DAISY for model \ref{Model3} without the initial conditions.}
    \label{fig:model3_daisy}
\end{figure}
Hence, \textit{Mathematica} and DAISY agree in parameter correlations. Next, we continue with  structural identifiability analysis of model \ref{Model3} in the case when the initial conditions are known. DAISY gives the following parameter correlations  (see Code \ref{fig:d_res_m31}):
$$ \D\frac{N}{\beta_A} = \D\frac{\hat{N}}{\hat{\beta}_A},\;\D\frac{\beta_I}{\beta_A} = \D\frac{\hat{\beta}_I}{\hat{\beta}_A},\rho = \hat{\rho},\; k = \hat{k},\; \gamma = \hat{\gamma} $$
and declares that the model \ref{Model3} is not structurally identifiable.  However, $N$ is the total population size and \ref{Model3} represents a closed population, the initial conditions determine the size of the population. Hence, the original model \ref{Model3} becomes identifiable when all the initial conditions are known. We summarize the structural identifiability analysis result when the initial conditions are known in the following Proposition \ref{m3pop}.

\begin{pro} \label{m3pop}
The model \ref{Model3} is structurally identifiable from the observations of new symptomatic infections if all the initial conditions are known.
\end{pro}

Recall that, model \eqref{Model3} is not structurally identifiable when the initial conditions are not known. In such a situation, when the model is not identifiable the next step is to reformulate the model and obtain a structurally identifiable version. To do so, we use the information regarding parameter correlations \eqref{m3res}.  We can obtain a structurally identifiable model by fixing the total population size, $N$, or the fraction of symptomatic individuals $\rho$ and by setting $\beta_A =\beta_I$. Thus we can modify the model \ref{Model3} as follows:

\begin{equation}\label{Model31}\tag{M$_3'$}
\textbf{Model 3' (M$_3'$):}
\begin{cases}
\displaystyle\frac{dS}{dt} = -\frac{\beta( A + I)S}{N}, \quad S(0) = S_0 \\[1.5ex]
\displaystyle \frac{dE}{dt} = \frac{\beta( A + I)S}{N} - kE,  \quad E(0) = E_0\\[1.5ex]
\displaystyle \frac{dI}{dt} = k\rho E-\gamma I, \quad I(0) = I_0 \\[1.5ex]
\displaystyle \frac{dA}{dt} =k(1-\rho) E -\gamma A, \quad A(0)=A_0 \\[1.5ex]
\displaystyle \frac{dR}{dt} = \gamma I+\gamma A, \quad R(0) = R_0. 
\end{cases}
\end{equation}

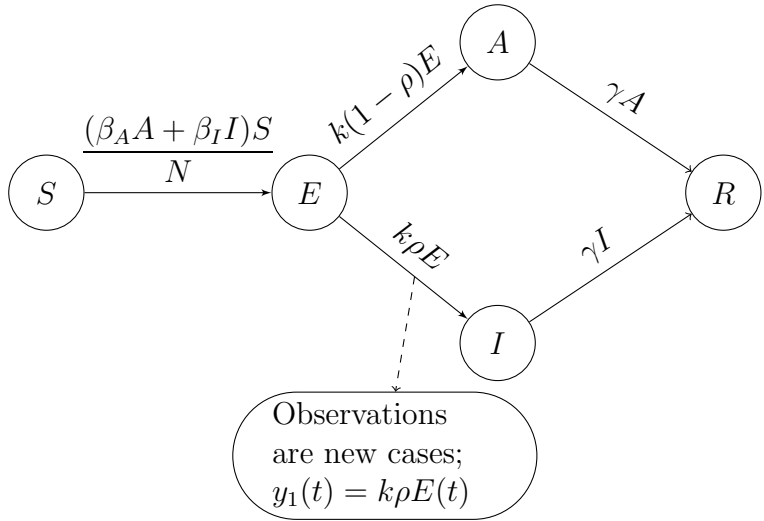
\begin{figure}
\caption{Compartmental diagram for Model \ref{Model3}. Circles show the epidemiological compartments for the different states of the system. Solid arrows indicate the transitions between compartments. The dashed arrows indicate the source of the observations, which are the newly infected symptomatic individuals. The structural identifiability analysis results are summarized when initial conditions are known and unknown.}\label{fig:model3}
\tikzstyle{line} = [draw, -latex']

\begin{tikzpicture} 
  \node [circle,draw,minimum size=1cm] (S) at (0,0) {$S$}; 
  \node [circle,draw,minimum size=1cm] (E) at (3.5,0) {$E$}; 
  \node [circle,draw,minimum size=1cm]  (I) at (6,-2) {$I$}; 
  \node [circle,draw,minimum size=1cm]  (A) at (6,2) {$A$}; 
  \node [circle,draw,minimum size=1cm]  (R) at (9,0) {$R$};
  \node [draw,rounded rectangle,text width=3cm] (O) at (4.5,-3.5) {Observations are new cases; $y_1(t) = k \rho E(t)$};
  \path [line]    (S) -- node [midway,above ] {$ \D \frac{(\beta_A A +\beta_I I)S}{N}$}  (E); 
  \draw [line]   (E) -- node [midway,above,sloped ] (c){$k \rho E $\\[.2em]} (I); 
  \draw [line]   (E) -- node [midway,above,sloped ] {$k (1-\rho) E $\\[.2em]} (A); 
  \draw [->]     (I) -- node [midway,above,sloped ] {$\gamma I $\\[.2em]}  (R); 
  \draw [->]     (A) -- node [midway,above,sloped ] {$\gamma A $\\[.2em]}  (R); 
  \draw[->, dashed] (c) --(O);
  \node [draw,text width=6cm] (idr) at (1,-7) {\textbf{Identifiability Results:}\\\textbf{Without Initial Conditions}\\ identifiable parameters: $k,\gamma$ \\ unidentifiable parameters: $\beta_A,\beta_I, N, \rho$\\
  parameter correlations:\\
 $ \beta_I \rho +\beta_A (1- \rho) = \hat\beta_I \hat\rho + \hat \beta_A (1-  \hat\rho)$,  $N \rho =\hat N \hat\rho$};
   \node [draw,text width=6cm] (idr) at (9,-7) {\textbf{Identifiability Results:}\\\textbf{With Initial Conditions}\\ identifiable parameters: $k,\gamma,\beta_A,\beta_I,N, \rho$ \\ };
\end{tikzpicture}
\end{figure}

\begin{pro}
When the total population size $N$ is known or the fraction of the symptomatic individuals, $\rho$ is known, the epidemic model \ref{Model31} is structurally identifiable from the observations of newly symptomatic infections.
\end{pro}
\begin{figure}
    \centering
    {\scriptsize
\begin{verbatim}

IDENTIFIABILITY WITH ALL INITIAL CONDITIONS (IC_)$

ic_ := {x1=x10,x2=x20,x3=x30,x4=x40,x5=x50}$

MODEL PARAMETER SOLUTIONS$

gi_ := {{n=(13*betaa)/2, betai=(3*betaa)/2 ,rho1=7, gamma1=11,k=5}}$

\end{verbatim}
  }
    \captionof{Code}{Parameter correlations derived by DAISY for model \ref{Model3} with known initial conditions.}
    \label{fig:d_res_m31}
\end{figure}
The enhanced compartmental diagram of model \ref{Model3} that includes the observed states of the system and the parameter correlations resulting from the structural identifiability analysis is given in Figure \ref{fig:model3}.

\subsection{SEIR model with disease-induced deaths (\ref{Model4})}

In the first three epidemic models (\ref{Model1}, \ref{Model2}, and \ref{Model3}), we considered diseases where infected individuals eventually recover from the infection. Next, we consider the epidemic model \ref{Model4} that includes disease-induced deaths \citep{brauer2019mathematical}, which is given by the following system of differential equations:

\begin{equation}\label{Model4}\tag{M$_4$}
\textbf{Model 4 (M$_4$):}
\begin{cases}
\displaystyle\frac{dS}{dt} = -\beta\frac{SI}{N(t)},  \quad S(0)=S_0\\[1.5ex]
\displaystyle \frac{dE}{dt} = \beta\frac{SI}{N(t)} - kE, \quad E(0)=E_0  \\[1.5ex]
\displaystyle \frac{dI}{dt} = k E-(\gamma+\delta)I, \quad I(0)=I_0 \\[1.5ex]
\displaystyle \frac{dR}{dt} =\gamma I, \quad R(0)=R_0 \\[1.5ex]
\displaystyle \frac{dD}{dt} = \delta I,  \quad D(0)=D_0.
\end{cases}
\end{equation}

Here, susceptible individuals move to exposed class at rate $\beta I(t)/N$, $k$ is the rate at which the exposed individuals become infectious, $\gamma$ is the recovery rate, and $\delta$ denotes the mortality rate of infectious individuals. Since the model includes disease-induced deaths, the total population $N(t)$ is not constant. Hence, $N$ is not a parameter in this model. The full parameter set of the model \ref{Model4} is given by  $\bfa p =\left(\beta,k,\gamma, \delta\right)$. For epidemics where disease-induced deaths occur, it is possible to observe incident deaths due to the disease. For this epidemic model \ref{Model4}, we will consider the structural identifiability of the model parameters under two different observation scenarios: (i) new infections are observed, (ii) new infections and new deaths are observed. \\

\underline{\textbf{(i) Number of new infected cases are observed:}}
First, we consider the scenario where the observations correspond to the number of newly infected cases. That is, $$y_1(t)= k E(t).$$ 
Using DAISY, we analyze the parameter identifiability results of the model. DAISY assigns 
$$ \hat{\bfa p} = \left\{ \hat{\beta}=2, \hat{k}=3, \hat{\gamma}=5, \hat{\delta}=7 \right\} $$ and  solve $c(\bfa p) = c(\hat{\bfa p})$ for these integers and yields,
$$\frac{\beta}{\delta}= \frac{2}{7},\quad \gamma+\delta=12, \quad k=3.$$
These results indicate the presence of parameter correlations. Specifically, we can identify parameter $k$, the ratio $\D\frac{\beta}{\delta}$ and the sum $\gamma+\delta.$ That is, any combination of parameters $\beta$ and $\delta$ that yields the same ratio $\beta/\delta$  or any combination of parameters $\delta$ and $\gamma$ that yields the same sum $\delta+\gamma$ will yield the same number of infected cases. We conclude that the model \ref{Model4} in its present form is not structurally identifiable from the observations. It only reveals the parameter $k$, from the newly infected cases.
\begin{pro}
The model \ref{Model4} is not structurally identifiable from the observation of the number of new infected cases.
\end{pro}
The parameter correlations derived by DAISY can be used to modify the model in order to make the parameters structurally identifiable.  Here, we fix $\delta,$ then the parameter correlations become,$$\beta=\frac{2}{7}\delta,\quad \gamma=12-\delta.$$ Since $\delta$ is known, we can identify $\gamma$ and $\beta.$
\begin{pro}
If $\delta$ is fixed, then the model \ref{Model4} becomes structurally identifiable from the observation of the number of newly infected cases.
\end{pro}

Alternatively, one can see that if $\gamma$ is fixed, then the model \ref{Model4} can become structurally identifiable from the number of new cases.

We note that DAISY did not produce any results for the case when all the initial conditions are known. This process was computationally too expensive for DAISY, which is likely due to the long characteristic set (see DAISY output file in our GitHub repository). We also note that it may be possible to analyze the input-output equation derived by DAISY using \textit{Mathematica} as in previous models. However, manually transferring this equation to \textit{Mathematica} increases the chances of introducing errors.\\

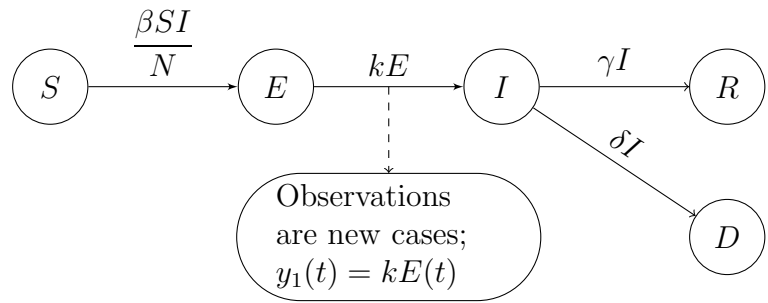
\begin{figure}
\caption{Compartmental diagram for Model \ref{Model4}. Circles show the epidemiological compartments for the different states of the system. Solid arrows indicate the transitions between compartments. The dashed arrows indicate the source of the observations. Structural identifiability analysis results are summarized when initial conditions are known and unknown.}\label{fig:model4}
\tikzstyle{line} = [draw, -latex']
\begin{tikzpicture} 
  \node [circle,draw,minimum size=1cm] (S) at (0,0) {$S$}; 
  \node [circle,draw,minimum size=1cm] (E) at (3,0) {$E$}; 
  \node [circle,draw,minimum size=1cm]  (I) at (6,0) {$I$}; 
  \node [circle,draw,minimum size=1cm]  (R) at (9,0) {$R$};
  
  \node [circle,draw,minimum size=1cm]  (D) at (9,-2) {$D$};
  \node [draw,rounded rectangle,text width=3cm] (O) at (4.5,-2) {Observations are new cases; $y_1(t) = kE(t)$};
  \path [line]    (S) -- node [midway,above ] {$ \D \frac{\beta S I}{N}$}  (E); 
  \draw [line]   (E) -- node [midway,above,sloped ] (c){$k E $\\[.2em]} (I); 
  \draw [->]     (I) -- node [midway,above,sloped ] {$\gamma I $\\[.2em]}  (R); 
   \draw [->]     (I) -- node [midway,above,sloped ] {$\delta I $\\[.2em]}  (D); 
    
  \draw[->, dashed] (c) --(O);
  \node [draw,text width=6cm] (idr) at (1,-7) {\textbf{Identifiability Results:\\ Without Initial Conditions}\\ identifiable parameters: $k$ \\ unidentifiable parameters: $\beta,\gamma,\delta$\\
  parameter correlations $\D\frac{\beta}{\delta} =\D\frac{\hat\beta}{\hat \delta} \quad \gamma+\delta = \hat \gamma +\hat \delta$};
  
  \node [draw,text width=6cm] (idr) at (9,-7) {\textbf{Identifiability Results:}\\\textbf{With Initial Conditions}\\ identifiable parameters: $k,\beta,\gamma,\delta$\\ };

\end{tikzpicture}
\end{figure}


\underline{\textbf{(ii) Number of new infected cases and deaths are observed:}}
Now, we investigate the scenario when two different observations are available for the same epidemic: The number of new cases $y_1(t) = k E(t)$, and the number of new deaths $y_2(t) = \delta I(t).$ We obtain the following set of input-output equations in DAISY,

\begin{equation}
\label{Model4_input_output}
\begin{aligned}
0&= y_2' - \delta y_1+y_2(\delta+\gamma)\\
    0&= y_1''' y_1' y_2^2+ y_1''' y_1 y_2^2 k+(y_1'')^2y_2^2\frac{(-\beta+2\delta)}{(\beta-\delta)}-y_1'' y_1'y_1 y_2 \frac{\delta(\beta+\delta)}{(\beta-\delta)}+ \\& y_1'' y_1' y_2^2\frac{(\beta\delta+\beta\gamma-\beta k+\delta^2+\delta\gamma+3\delta k)}{(\beta-\delta)}-y_1'' y_1^2 y_2 \frac{\delta k(\beta+\delta)}{(\beta-\delta)}+\\& y_1'' y_1 y_2^2 \frac{k(\beta\delta+\beta\gamma+ \beta k+\delta^2+\delta\gamma-\delta k)}{(\beta-\delta)}- y_1'^3 y_2 \delta+y_1'^2 y_1^2 \frac{\delta^2(2\beta-\delta)}{(\beta-\delta)}+\\& y_1'^2 y_1 y_2 \frac{\delta(-3\beta\delta-3\beta\gamma- 3\beta k+\delta^2+\delta\gamma+\delta k)}{(\beta-\delta)}+\\&y_1'^2 y_2^2\frac{(\beta\delta^2+2\beta\delta\gamma+\beta\delta k+\beta\gamma^2+\beta\gamma k-\beta k^2+\delta^2 k+\delta\gamma k+2 \delta k^2)}{(\beta-\delta)}+\\& 2y_1' y_1^3\frac{\delta^2 k(2\beta-\delta)}{(\beta-\delta)}+ 2y_1' y_1^2 y_2 \frac{\delta(-3\beta\delta-3\beta\gamma-\beta k+\delta^2+\delta\gamma)}{(\beta-\delta)}+\\&y_1' y_1 y_2^2 \frac{k(2\beta\delta^2+4\beta\delta\gamma+ \beta\delta k+2\beta\gamma^2+\beta\gamma k+\delta^2 k+\delta\gamma k)}{(\beta-\delta)}+ y_1^4 \frac{\delta^2 k^2(2\beta-\delta)}{(\beta-\delta)}+\\&y_1^3 y_2 \frac{\delta k^2(-3\beta\delta-3\beta\gamma+\delta^2+\delta\gamma)}{(\beta-\delta)}+\\& y_1^2 y_2^2 \frac{\beta k^2(\delta^2+2\delta\gamma+\gamma^2)}{(\beta-\delta)}
\end{aligned}
\end{equation}
Since we have two sets of observations, we obtain two monic differential polynomials involving both observations. Thus, the input-output equations \eqref{Model4_input_output} of the model \ref{Model4} involves both $y_1$ and $y_2$ together with the model parameters. Following the definition of structural identifiability, we suppose the parameter set  $\hat{ \bfa p}=(\hat{k},\hat{\beta},\hat{\gamma},\hat{\delta})$ produces the same observations $y_1$ and $y_2$ as the parameter $\bfa p$.  Then setting $c(\bfa p) = c(\hat{\bfa p})$, we obtain the following set of solutions,
\begin{equation*}
\left\{k=\hat{k},\quad \beta=\hat{\beta},\quad \gamma=\hat{\gamma},\quad \delta=\hat{\delta}\right\}
\end{equation*}
 This implies that the model parameters for \ref{Model4} are structurally identifiable from the observations of new cases $y_1(t) = k E(t)$, and new deaths $y_2(t) = \delta I(t).$  Note that this result does not consider the initial conditions. Thus, the model \ref{Model4} is structurally identifiable whether the initial conditions are known or not. We state the following Proposition \ref{pro_4}:
 
 \begin{pro} \label{pro_4}
 The model \ref{Model4} is structurally identifiable from the observation of newly infected cases, $y_1(t) = k E(t)$ and deaths, $y_2(t) = \delta I(t)$.
 \end{pro}
 
In summary, when only the new cases are observed, it was not possible to identify all parameters without modifying the model according to the parameter correlations provided by DAISY. With additional observations corresponding to the number of new deaths, the original model becomes structurally identifiable without needing any modifications.
 
 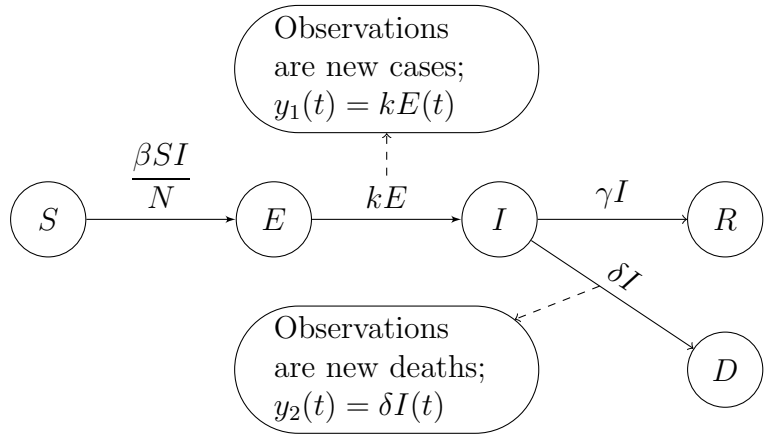
\begin{figure}
\caption{Compartmental diagram of model \ref{Model4}. Circles show the epidemiological compartments for the different states of the system. Solid arrows indicate the transitions between compartments. The dashed arrows indicate the source of the observations, which are the newly infected cases and deaths. Results of the structural identifiability analysis are also summarized when the initial conditions are not known.}

\label{fig:model4b}
\tikzstyle{line} = [draw, -latex']
\begin{tikzpicture} 
  \node [circle,draw,minimum size=1cm] (S) at (0,0) {$S$}; 
  \node [circle,draw,minimum size=1cm] (E) at (3,0) {$E$}; 
  \node [circle,draw,minimum size=1cm]  (I) at (6,0) {$I$}; 
  \node [circle,draw,minimum size=1cm]  (R) at (9,0) {$R$};
  \node [circle,draw,minimum size=1cm]  (D) at (9,-2) {$D$};
  \node [draw,rounded rectangle,text width=3cm] (O) at (4.5,-2) {Observations are new deaths; $y_2(t) = \delta I(t)$};
   \node [draw,rounded rectangle,text width=3cm] (m) at (4.5,2) {Observations are new cases; $y_1(t) = kE(t)$};
  \path [line]    (S) -- node [midway,above ] {$ \D \frac{\beta S I}{N}$}  (E); 
  \draw [line]   (E) -- node [midway,above,sloped ] (c){$k E $\\[.2em]} (I); 
  \draw [->]     (I) -- node [midway,above,sloped ] {$\gamma I $\\[.2em]}  (R); 
   \draw [->]     (I) -- node [midway,above,sloped ] (f){$\delta I $\\[.2em]}  (D); 
    
   \draw[->, dashed] (f) --(O);
   \draw[->, dashed] (c) --(m);
  \node [draw,text width=8cm] (idr) at (3,-5.5) {\textbf{Identifiability Results:\\ Without  Initial Conditions}\\ identifiable parameters: $k,\gamma,\beta,\delta$\\ unidentifiable parameters: None};

\end{tikzpicture}
\end{figure}
The enhanced compartmental diagram and results of the structural identifiability analysis of model  \ref{Model4} when including the new cases, and new cases as well as new deaths are given in Figure \ref{fig:model4} and Figure \ref{fig:model4b}.


\subsection{Simple vector-borne disease model (\ref{Model5})}

Next, we consider epidemic models designed to study the transmission dynamics of vector-borne diseases. The next model \ref{Model5} is a simple vector-borne model given by the following set of differential equations:

\begin{equation}\label{Model5}\tag{M$_5$}
	\textbf{Model 5 (M$_5$):}
	\begin{cases}
		\displaystyle \frac{dS_{\upsilon}}{dt} = \Lambda_{\upsilon} - \frac{\beta S_{\upsilon}I}{N} -\mu_{\upsilon} S_\upsilon, \quad S_\upsilon(0) = S_{\upsilon0}\\[1.5ex]
		\displaystyle \frac{dI_{\upsilon}}{dt} 
= \frac{\beta S_{\upsilon}I}{N}-\mu_{\upsilon} I_{\upsilon}, \quad I_{\upsilon}(0) =I_{\upsilon0}  \\[1.5ex]
		\displaystyle \frac{dS}{dt} 
=  -\frac{\beta_\upsilon S I_{\upsilon}}{N}, \quad S(0) = S_0 \\[1.5ex]
		\displaystyle \frac{dI}{dt} 
=  \frac{\beta_\upsilon S I_{\upsilon}}{N} - \gamma I, \quad I(0)=I_0 \\[1.5ex]
		\displaystyle \frac{dR}{dt} 
=  \gamma I, \quad R(0) = R_0.
	\end{cases}
\end{equation}

This model is equipped with five epidemiological states over time: the number of susceptible mosquitoes ($S_{\upsilon}(t)$),  the number of infected mosquitoes ($I_{\upsilon}(t)$), the number of susceptible individuals ($S(t)$), the number of infected individuals ($I(t)$), and the number of recovered individuals ($R(t)$). The mosquito birth rate is a constant given by ${\Lambda}_{\upsilon}$ whereas the mosquito per-capita death rate is given by ${\mu}_{\upsilon}$. Susceptible mosquitoes are infected by biting infectious humans at the rate $\beta I(t)/N$ whereas susceptible humans are infected by infectious mosquitoes at the rate ${\beta_\upsilon} I_{\upsilon}(t)/N$. Like before, an infected individual recovers at rate ${\gamma}$.
 In this model, the total human population $N$ remains the same throughout the epidemic, hence  $N$ = $S_0$ +$I_0$ +$R_0$. The full parameter set of the model is given by $\bfa p = \left(\Lambda_\upsilon, \mu_\upsilon, \beta_\upsilon, N, \beta, \gamma\right)$. 
First, we used the number of new cases as the observations for the vector-borne model \ref{Model5}. However, DAISY could not finish the computations to produce the input-output equations. Hence, we tried changing the observations to the cumulative number of new cases, denoted by $C(t)$. In terms of the model variables we define the cumulative number of cases as; 
\begin{equation}\label{model5_data}
	y_1(t)= C(t)=\int_{0}^{t}\frac{\beta_\upsilon S I_{\upsilon}}{N}=S(0)-S(t).
\end{equation}
So, if we know the initial number of susceptibles, $S_0$, the cumulative number of infections is equivalent to observing $S(t)$. Next, we investigate whether the epidemic model \ref{Model5} can reveal its epidemiological parameters from the cumulative number of new infections. 
DAISY assigns the following integers to the model parameters, 
$$\hat{\bfa p} = \left\{\hat{\Lambda}_\upsilon = 2,\; \hat{\beta} =3, \; \hat{\mu}_\upsilon=5,\; \hat{\beta}_\upsilon = 7,\hat{\gamma} = 11,\hat{N} = 13\right\}$$ and then solves 
$c(\bfa p) = c(\hat{\bfa p})$ and gives the following correlations between the parameters, which the code and the results are available at our GitHub site.
\begin{equation}\label{D_res_m5}
\D\frac{\beta_\upsilon\Lambda_\upsilon}{N}= \D\frac{14}{13}  = \D\frac{\hat{\beta}_\upsilon\hat{\Lambda}_\upsilon}{\hat{N}}, \quad
\D\frac{\beta}{N}=\D\frac{3}{13} =\D\frac{\hat\beta}{\hat N}, \quad 
\gamma=11 =\hat{\gamma}, \quad \text{and} \quad \mu_\upsilon=5 = \hat{\mu}_\upsilon.
\end{equation}
 DAISY result \eqref{D_res_m5} indicates that only the parameters $\gamma$ and $\mu_\upsilon$ are identifiable, and the parameters $\beta, \Lambda_\upsilon, N$ and  $\beta_\upsilon$ are not identifiable. We can only identify the ratios,
$$ \D\frac{\beta_\upsilon\Lambda_\upsilon}{N} \quad \text{and} \quad \D\frac{\beta}{N}.$$
Hence, we can state the first structural identifiability result regarding the vector-borne model \ref{Model5} in Proposition \ref{prop1_m5}.

\begin{pro} \label{prop1_m5}
The model \ref{Model5} is not structurally identifiable from the observation of cumulative incidences. Only the parameters $\gamma, \mu_\upsilon$  can be identified. Parameters  $\beta, \Lambda_\upsilon, N$ and  $\beta_\upsilon$  can not be identified since they are correlated to each other by
 $ \D\frac{\beta_\upsilon\Lambda_\upsilon}{N}$ and  $\D\frac{\beta}{N}$. 
\end{pro}
According to the parameter correlations above \eqref{D_res_m5}, $\beta_\upsilon$ can only be identified if $N$ and $\Lambda_\upsilon$ are known. Moreover, $\beta$ can only be identified when $N$ is known. Hence, we can define the following result.

\begin{pro} \label{prop2_m5}
When the total human population $N$ and the recruitment rate $\Lambda_\upsilon$ of the vector population are known, the model \ref{Model5} becomes structurally identifiable from the observation of cumulative incidences. 
\end{pro}

Next we carry out the structural identifiability analysis of the vector-borne model \ref{Model5} when all the initial conditions are known. With known initial conditions DAISY solves $c(\bfa p) = c(\hat{\bfa p})$ together with the characteristic set evaluated at time $t=0$ and provides the following solution and declares that the model is not identifiable.
$$\Lambda_\upsilon =2,\;
\D\frac{\beta}{N}=\frac{3}{13}, \;
\D\frac{\beta_\upsilon}{N}= \frac{7}{13},\;
\gamma=11, \; \text{and} \; \mu_\upsilon=5. $$ 

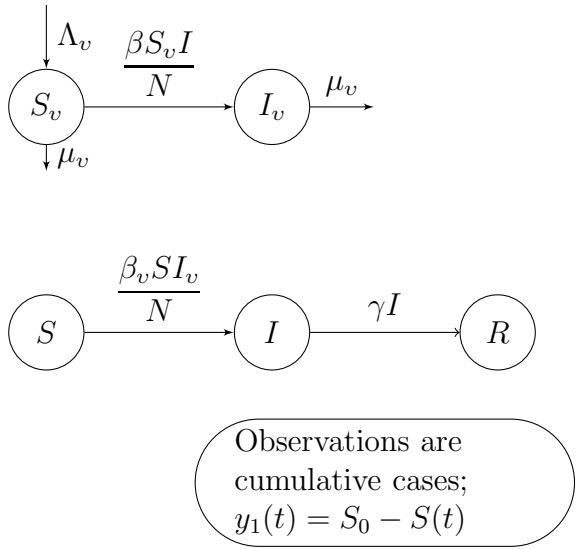
\begin{figure}
\caption{Compartmental diagram for Model \ref{Model5}. Circles show the epidemiological compartments for the different states of the system. Solid arrows indicate the transitions between compartments. The dashed arrows indicate the source of the observations, which are the cumulative number of new cases. Structural identifiability analysis results are summarized when initial conditions are known and unknown.}\label{fig:model5}
\tikzstyle{line} = [draw, -latex']
\begin{tikzpicture} 
  \node [circle,draw,minimum size=1cm] (Sv) at (0,3) {$S_\upsilon$}; 
  \node [circle,draw,minimum size=1cm]  (Iv) at (3,3) {$I_\upsilon$}; 
  \node [circle,draw,minimum size=1cm] (S) at (0,0) {$S$}; 
  \node [circle,draw,minimum size=1cm]  (I) at (3,0) {$I$}; 
  \node [circle,draw,minimum size=1cm]  (R) at (6,0) {$R$};
  \node (lv) at (0,4.5) {};
  \node (mvs) at (0,2) {};
  \node (mv) at (4.5,3) {};
  \node [draw,rounded rectangle,text width=4cm] (O) at (4.5,-2) {Observations are cumulative cases; $y_1(t) = S_0- S(t)$};
    \path [line]    (Sv) -- node [midway,above ] {$ \D \frac{\beta S_\upsilon I}{N}$}  (Iv); 
  \path [line]    (S) -- node [midway,above ] {$ \D \frac{\beta_\upsilon S I_\upsilon}{N}$}  (I); 
  \draw [->]     (I) -- node [midway,above,sloped ] {$\gamma I $\\[.2em]}  (R); 
  \path [line] (lv)-- node [midway,right] {$\Lambda_\upsilon$} (Sv);
  \path [line] (Sv)-- node [midway,right] {$\mu_\upsilon$} (mvs);
  \path [line] (Iv)-- node [midway,above] {$\mu_\upsilon$} (mv);
  \node [draw,text width=6cm] (idr) at (1,-5.5) {\textbf{Identifiability Results:}\\\textbf{Without Initial Conditions}\\ identifiable parameters: $\mu_\upsilon,\gamma$ \\ unidentifiable parameters: $\beta,\beta_\upsilon, \Lambda_\upsilon, N$\\
  parameter correlations $\D\frac{\beta}{N} =\D\frac{\hat\beta}{\hat N} $ and $\D\frac{\beta_\upsilon\Lambda_\upsilon}{N} = \D\frac{\hat{\beta}_\upsilon\hat{\Lambda}_\upsilon}{\hat{N}}$};
   \node [draw,text width=6cm] (idr) at (8,-5.5) {\textbf{Identifiability Results:}\\\textbf{With Initial Conditions}\\ identifiable parameters: $\beta,\beta_\upsilon,\Lambda_\upsilon, \mu_\upsilon,\gamma,N$ \\ };
\end{tikzpicture}
\end{figure}

Basically, known initial conditions only revealed  mosquito birth rate, $\Lambda_\upsilon$, and the parameter correlations remained the same.  On the other hand,  $N$ = $S_0$ +$I_0$ +$R_0$, since all the initial conditions are known, $N$ will be identifiable, which, in turn, makes $\beta$ and $\beta_\upsilon$ identifiable. We conclude that the model is structurally identifiable when the initial conditions are known and state the result in the following proposition.
\begin{pro}
The model \ref{Model5} is structurally identifiable from the observation of cumulative cases when all the initial conditions are known.
\end{pro}
The enhanced compartmental diagram of model \ref{Model5} that includes the observed states of the system and the parameter correlations resulting from the structural identifiability analysis is given in Figure \ref{fig:model5}.

\subsection{Vector-borne disease model with asymptomatic infections (\ref{Model6})}

We proceed to conduct the structural identifiability analysis of the vector-borne model, \ref{Model6}, which incorporates an asymptomatic class to the previous vector-borne model \ref{Model5} and is given by:

  \begin{equation}\label{Model6}\tag{M$_6$}
	\textbf{Model 6 (M$_6$):}
	\begin{cases}
	\displaystyle \frac{dS_{\upsilon}}{dt} 
= \Lambda_{\upsilon} - \frac{\beta_{I} S_{\upsilon}I+\beta_{A} S_{\upsilon} A}{N} -\mu_{\upsilon}S_{\upsilon}, \quad S_\upsilon(0) = S_{\upsilon0}\\[1.5ex]
	\displaystyle \frac{dI_{\upsilon}}{dt} 
= \frac{\beta_{I} S_{\upsilon}I+\beta_{A} S_{\upsilon} A}{N}-\mu_{\upsilon} I_{\upsilon}, \quad I_\upsilon(0) = I_{\upsilon0}\\[1.5ex]
	\displaystyle \frac{dS}{dt} 
= - \frac{\beta_{\upsilon} I_{\upsilon}S}{N}, \quad S(0) = S_0\\[1.5ex]
	\displaystyle \frac{dI}{dt} 
= \frac{ \rho \beta_{\upsilon} I_{\upsilon}S}{N}-\gamma I, \quad I(0) = I_0\\[1.5ex]
	\displaystyle \frac{dA}{dt} 
=   \frac{(1-\rho)\beta_{\upsilon}I_{\upsilon}S }{N}  -\gamma A, \quad A(0) = A_0\\[1.5ex]
	\displaystyle \frac{dR}{dt} 
=  \gamma I +\gamma A, \quad R(0) = R_0.\\[1.5ex]

	\end{cases}
\end{equation}

Therefore, this vector-borne model \ref{Model6} has six epidemiological states: the number of susceptible mosquitoes ($S_{\upsilon}(t)$),  the number of infected mosquitoes ($I_{\upsilon}(t)$), the number of susceptible individuals ($S(t)$), the number of infected individuals ($I(t)$), the number of asymptomatic individuals ($A(t)$) and the number of recovered individuals ($R(t)$). As before, the mosquito birth rate is a constant given by ${\Lambda}_{\upsilon}$ whereas the mosquito per-capita death rate is given by ${\mu}_{\upsilon}$. Likewise, ${\gamma}$ is the recovery rate of the human population. Since there are two infectious classes, there are two routes of transmission. The first route is when a susceptible mosquito bites a symptomatic infectious individual and the other is when a susceptible mosquito bites an asymptomatic infectious individual. We model the infected incidence in the mosquito population with the following term, 
$$ \D\frac{\beta_{I} S_{\upsilon}I+\beta_{A} S_{\upsilon} A}{N}$$

Susceptible mosquitoes get infected from symptomatic individuals at the rate ${\beta_I} I/N$ and from infected asymptomatic individuals at the rate ${\beta_A A/N}$. A susceptible human becomes infected at rate ${\beta_\upsilon} I_{\upsilon}/N$ from infectious mosquitoes.  Parameter ${\rho}$ models the proportion of new infections that show symptoms and move to the symptomatic class, and the remaining fraction ($1-{\rho}$) of the new infections move to the asymptomatic class. The total human population, $N$, remains the same throughout the epidemic, hence  $N = S_0 +I_0+ A_0 +R_0$. The full parameter set of the model \ref{Model6} that we would like to identify given the observations is $\bfa p = \left(N, \Lambda_\upsilon, \mu_\upsilon, \beta_\upsilon, \beta_A,\beta_I, \gamma, \rho \right)$. 
We use the cumulative number of symptomatic cases as the observations for this vector-borne model \ref{Model6}. 
\begin{equation}\label{model6_data}
	y_1(t)= C(t)=\rho\int_{0}^{t} \frac {S I_\upsilon \beta_\upsilon}{N} = \rho(S(0)-S(t)). 
\end{equation}

As seen in  \eqref{model6_data}, the cumulative incidence is equivalent to $\rho(S(0)-S(t))$.  We first investigate whether the epidemic model \ref{Model6} can reveal its epidemiological parameters, $\bfa p$
from the observations alone.
DAISY first assigns the following integers for the model parameters (see Code \ref{fig:d_res_m6}), 
$$ \hat{\bfa p} = \left\{ \hat{\beta}_I=2,\; \hat{\beta}_A=3, \; \hat{\mu}_\upsilon=5,\; \hat{\beta}_\upsilon =7, \; \hat{\rho} =11,\;\hat{\gamma}=13, \;\hat{\Lambda}_\upsilon=17,\; \hat{N}=19\right \}$$
and then solves $ c(\bfa p) = c(\hat{\bfa p})$ and gives the following results: 

\begin{figure}
    \centering
    {\scriptsize
\begin{verbatim}
PARAMETER VALUES$

b2_ := {betai=2,betaa=3,muv=5,betav=7,rho1=11,gamma1=13,lambdaupsilon=17,n=19}$

MODEL PARAMETER SOLUTION(S)$

g_ := {{betaa=(209*betai + 8*n)/190,
betav=(119*n)/(19*lambdaupsilon),
gamma1=13,
muv=5,
rho1=11}}$

MODEL NON IDENTIFIABLE$

IDENTIFIABILITY WITH ALL INITIAL CONDITIONS (IC_)$

ic_ := {x1=x10,x2=x20,x3=x30,x4=x40,x5=x50,x6=x60}$

MODEL PARAMETER SOLUTIONS$

gi_ := {{lambdaupsilon=17,
betai=(2*n)/19,
betaa=(3*n)/19,
betav=(7*n)/19,
gamma1=13,
muv=5,
rho1=11}}$


\end{verbatim}
  }
    \captionof{Code}{Parameter correlations derived by DAISY for model \ref{Model6}.}
    \label{fig:d_res_m6}
\end{figure}

\begin{equation}\label{d_res_m6}
209 \beta_I  + 8 N  = 190 \beta_A , \quad  
\D\frac{\beta_{\upsilon} \Lambda_{\upsilon}}{N} =\frac{119}{19}, \quad
\rho = 11,\quad
\gamma = 13, \quad
\mu_{\upsilon} = 5
\end{equation}

Please note that the first correlation in \eqref{d_res_m6}, $209 \beta_I  + 8 N  = 190 \beta_A $,  is the same correlation \eqref{same_corr} that we observed in model \ref{Model3} with an asymptomatic class. Namely, 
$$ \D\frac{ \beta_I \rho +\beta_A (1- \rho)}{N \rho} =  \D\frac{\hat\beta_I \hat\rho + \hat \beta_A (1-  \hat\rho)}{\hat N \hat\rho}$$ 
and $\rho =\hat{\rho} = 11.$ Indeed, we can get the same answer by substituting the integers for the parameters assigned by DAISY. The corresponding DAISY input code and results are available from our GitHub site.
The DAISY result \eqref{d_res_m6} indicates that the parameters $\gamma$ and $\mu_\upsilon$ are identifiable, and that the following correlations exist involving other parameters:
$$ \D\frac{\beta_{\upsilon} \Lambda_{\upsilon}}{N} =\D\frac{\hat{\beta}_{\upsilon} \hat{\Lambda}_{\upsilon}}{\hat N} \quad \text{and} \quad  \D\frac{ \beta_I \rho +\beta_A (1- \rho)}{N \rho} =  \D\frac{\hat\beta_I \hat\rho + \hat \beta_A (1-  \hat\rho)}{\hat N \hat\rho} $$ 
 This means, $\beta_\upsilon$ can be identified only if $N$ and $\Lambda_\upsilon$ are known and $\beta_I$ can be identified only when $\beta_A$, $N$  and $\rho$ are known.
\begin{pro}
The model \ref{Model6} is not structurally identifiable from the observation of cumulative number of new infections. 
\end{pro}

Next, we conduct the structural identifiability analysis when all the initial conditions are known. DAISY solves $c(\bfa p) = c(\hat{\bfa p})$ together with all the characteristic set evaluated at time $t=0$ gives the following solutions
$$ \Lambda_\upsilon =17 = \hat{\Lambda}_\upsilon,\quad 
\gamma=13 =\hat{\gamma},\quad  
\rho=11 = \hat{\rho}, \quad \text{and} \quad 
\mu_\upsilon=5 = \hat{\mu}_\upsilon$$
together with the following correlations
$$
\D\frac{\beta_\upsilon}{N}=\frac{7}{19} = \D\frac{\hat{\beta}_\upsilon}{\hat N},\quad
\D\frac{\beta_A}{N}= \frac{3 }{19} = \D\frac{\hat{\beta}_A}{\hat N},\quad
\D\frac{\beta_I}{N}=\frac{ 2 }{19} = \D\frac{\hat{\beta}_I}{\hat N},$$. 

\begin{figure}
\caption{Compartmental diagram for Model \ref{Model6}. Circles show the epidemiological compartments for the different states of the system. Solid arrows indicate the transitions between compartments. The dashed arrows indicate the source of the observations, which are the cumulative number of new cases. Structural identifiability analysis results are summarized for the scenario when the initial conditions are known and unknown.}\label{fig:model6}
\tikzstyle{line} = [draw, -latex']
\begin{tikzpicture} 
  \node [circle,draw,minimum size=1cm] (Sv) at (0,3) {$S_\upsilon$}; 
  \node [circle,draw,minimum size=1cm]  (Iv) at (5,3) {$I_\upsilon$}; 
  \node [circle,draw,minimum size=1cm] (S) at (0,0) {$S$}; 
  \node [circle,draw,minimum size=1cm]  (I) at (4.5,1) {$I$}; 
   \node [circle,draw,minimum size=1cm]  (A) at (4.5,-1) {$A$}; 
  \node [circle,draw,minimum size=1cm]  (R) at (8.5,0) {$R$};
  \node (lv) at (0,4.5) {};
  \node (mvs) at (0,2) {};
  \node (mv) at (7,3) {};
  \node [draw,rounded rectangle,text width=4cm] (O) at (4.5,-2.5) {Observations are cumulative cases; $y_1(t) = \rho(S_0- S(t))$};
    \path [line]    (Sv) -- node [midway,above ] {$\D\frac{\beta_{I} S_{\upsilon}I+\beta_{A} S_{\upsilon} A}{N}$}  (Iv); 
  \path [line]    (S) -- node [midway,above,sloped ] {$ \rho\D \frac{\beta_\upsilon S I_\upsilon}{N}$}  (I); 
   \path [line]    (S) -- node [midway,below, sloped] {$ (1-\rho)\D \frac{\beta_\upsilon S I_\upsilon}{N}$}  (A); 
  \draw [->]     (I) -- node [midway,above,sloped ] {$\gamma I $\\[.2em]}  (R); 
    \draw [->]     (A) -- node [midway,above,sloped ] {$\gamma A $\\[.2em]}  (R); 
  \path [line] (lv)-- node [midway,right] {$\Lambda_\upsilon$} (Sv);
  \path [line] (Sv)-- node [midway,right] {$\mu_\upsilon$} (mvs);
  \path [line] (Iv)-- node [midway,above] {$\mu_\upsilon$} (mv);
  \node [draw,text width=7cm] (idr) at (1,-6) {\textbf{Identifiability Results:}\\\textbf{Without Initial Conditions}\\ identifiable parameters: $\mu_\upsilon,\gamma,\rho$ \\ unidentifiable parameters: $\beta_A, \beta_I,\beta_\upsilon, \Lambda_\upsilon, N$\\
  parameter correlations  $\D\frac{\beta_\upsilon\Lambda_\upsilon}{N} = \D\frac{\hat{\beta}_\upsilon\hat{\Lambda}_\upsilon}{\hat{N}}$, 
  $\D\frac{ \beta_I \rho +\beta_A (1- \rho)}{N \rho} =  \D\frac{\hat\beta_I \hat\rho + \hat \beta_A (1-  \hat\rho)}{\hat N \hat\rho}$};
   \node [draw,text width=6cm] (idr) at (8,-5.5) {\textbf{Identifiability Results:}\\\textbf{With Initial Conditions}\\ identifiable parameters: $\beta_A,\beta_I,\beta_\upsilon,\Lambda_\upsilon, \mu_\upsilon,\gamma,\rho,N$ \\ };
\end{tikzpicture}
\end{figure}
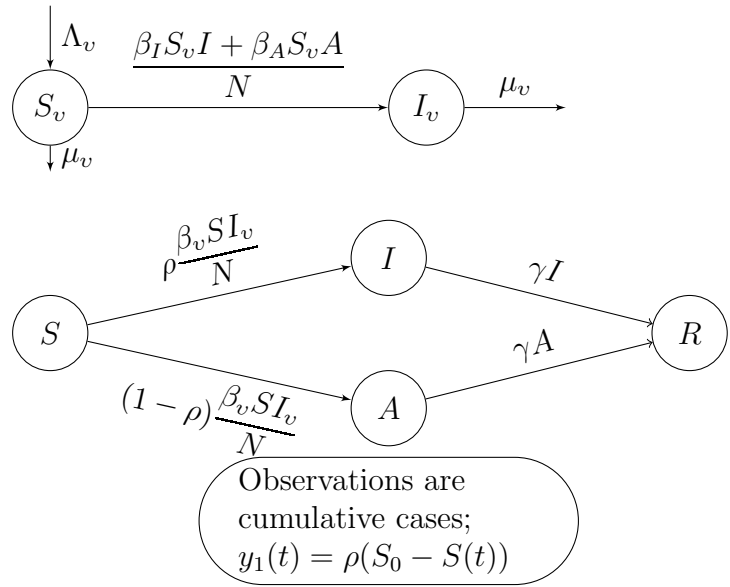

These correlations indicate that when we fix the total population size, $N$
the parameters $\beta_A$, $\beta_\upsilon$ and  $\beta_I$ will be identified.
Since, $N = S_0 +I_0+ A_0 +R_0$, we conclude that the model is structurally identifiable when the initial conditions are known. 
\begin{pro}
The model \ref{Model6} is structurally identifiable from the observation of cumulative cases when all the initial conditions are known.
\end{pro}
The enhanced compartmental diagram of model \ref{Model6} that includes the observed states of the system and the parameter correlations resulting from the structural identifiability analysis is given in Figure \ref{fig:model6}.

\subsection{Ebola model (\ref{Model7})}

This epidemic model is motivated by the transmission dynamics of Ebola, where transmission may be amplified in health-care settings and during funerals of Ebola deceased individuals \citep{legrand2007understanding}. The model is given the following system of differential equations:

\begin{equation}\label{Model7}\tag{M$_7$}
\textbf{Model 7 (M$_7$):}
\begin{cases}
\displaystyle\frac{dS}{dt}= \displaystyle\frac{-(\beta_{I} I + \beta_{H} H+\beta_{D} D) S}{N}, \quad S(0) = S_0\\[1.5ex]
 \displaystyle\frac{dE}{dt}=  \displaystyle\frac{(\beta_{I} I + \beta_{H} H+\beta_{D} D) S}{N}-kE, \quad E(0) = E_0\\[1.5ex]
\displaystyle\frac{dI}{dt}=kE-(\alpha+\gamma_{I}+\delta_{I})I, \quad I(0) = I_0\\[1.5ex]
\displaystyle\frac{dH}{dt}=  \alpha I -(\gamma_{H}+\delta_{H}) H,\quad H(0) = H_0\\[1.5ex]  
\displaystyle\frac{dR}{dt}=  \gamma_I I +\gamma_{H} H, \quad R(0) = R_0\\[1.5ex]
 \displaystyle\frac{dD}{dt}=  \delta_{I} I+\delta_{H} H, \quad D(0) = D_0.\\[1.5ex]
 \end{cases}
\end{equation}

It keeps track of six epidemiological states: The number of susceptible individuals $(S(t))$, the number of latent individuals $(E(t))$, the number of infected individuals in the community $(I(t))$, the number of hospitalized individuals $(H(t))$, the number of recovered individuals $(R(t))$ and the number of disease-induced deaths $(D(t)).$ Here, susceptible individuals can become exposed to Ebola virus by close contact with infected individuals in the community at the rate $\beta_I I(t)/N$, with hospitalized individuals at the rate $\beta_H H(t)/N$ or Ebola-deceased individuals at rate $\beta_D D(t)/N$. An exposed individual becomes infectious with Ebola after an average latent period of $1/k$ days. Infectious individuals in the community recover at rate $\gamma_I$  or die from the disease at rate $\delta_I$. A hospitalized individual can either recover at rate $\gamma_H$ or die at rate $\delta_H.$  The set of model parameters is given by $\bfa p=(\beta_I,\beta_H,\beta_D,k,\alpha,\delta_I,\delta_H,\gamma_H,\gamma_I)$. We will explore the structural identifiability of the model \ref{Model7} under three different observation situations: (i) number of new infections are observed, (ii) number of new infections and hospitalizations are observed, (iii) number of new infections, hospitalizations and deaths are observed.

 \underline{\textbf{(i) Number of new infected cases are observed}}: We will first examine the case when the observed data corresponds to the number of new infections, that is  $y_1(t) = kE(t).$ Using DAISY, we obtain the following input-output equation:
\begin{equation}\label{io_m71}
0=y_1'' + y_1'\frac{(\alpha\beta_H + \beta_D\delta_I + \beta_I\delta_H + \beta_I\gamma_H)}{\beta_I} + y_1\frac{\beta_D}{\beta_I} (
\alpha \delta_H + \delta_H\delta_I + \delta_I \gamma_H)
\end{equation}

Because parameters $k$ and $\gamma_I$ do not appear in the input-output equation \eqref{io_m71}, we conclude that these parameters are not structurally identifiable from the observations of new Ebola cases alone. Following the definition of structural identifiability, we first set  $c(\bfa p) = c(\bfa {\hat{p}})$, that is 
\begin{equation}\label{cp1} \D\frac{(\alpha\beta_H + \beta_D\delta_I + \beta_I\delta_H + \beta_I\gamma_H)}{\beta_I} = \D\frac{(\hat\alpha \hat{\beta}_H + \hat{\beta}_D \hat{\delta}_I + \hat{\beta}_I \hat{\delta}_H + \hat{\beta}_I \hat{\gamma}_H)}{\hat{\beta}_I}
\end{equation}
 and 
\begin{equation}\label{cp2}  \D\frac{\beta_D}{\beta_I} (
\alpha \delta_H + \delta_H\delta_I + \delta_I \gamma_H) = \D\frac{\hat{\beta}_D}{\hat{\beta}_I} (
\hat{\alpha} \hat{\delta}_H + \hat{\delta}_H \hat{\delta}_I + \hat{\delta_I}  \hat{\gamma}_H)
\end{equation}

\begin{figure}
\caption{Compartmental diagram for Model \ref{Model7}.Circles show the epidemiological compartments for the different states of the system. Solid arrows indicate the transitions between compartments. The dashed arrows indicate the source of the observations, which are the new infections. Structural identifiability analysis results are summarized for the scenario when initial conditions are known and unknown.}\label{fig:model7a}
\tikzstyle{line} = [draw, -latex']
\begin{tikzpicture} 
  
  \node [circle,draw,minimum size=1cm] (S) at (0,0) {$S$}; 
  \node [circle,draw,minimum size=1cm]  (E) at (4,0) {$E$}; 
  \node [circle,draw,minimum size=1cm]  (I) at (7,0) {$I$}; 
  \node [circle,draw,minimum size=1cm]  (R) at (10,0) {$R$}; 
   \node [circle,draw,minimum size=1cm]  (H) at (7,-2) {$H$}; 
  \node [circle,draw,minimum size=1cm]  (D) at (10,-2) {$D$};
  
 \node [draw,rounded rectangle,text width=4cm] (O) at (3.5,-3.5) {Observations are new infections; $y_1(t) = kE$};

 \node (z) at (5.5,0) {};
 \draw[->, dashed] (z) --(O);
    
 \path [line]    (S) -- node [midway,above ] {$\frac{S(\beta_I I+\beta_H H+\beta_D D)}{N}$}  (E); 
 \draw [->]     (E) -- node [midway,above,sloped ] {$kE$\\[.2em]}  (I);
 
 \draw [->]     (I) -- node [midway,above,sloped ] {$\gamma_I I$\\[.2em]}  (R);
 
 \draw [->]     (I) -- node [midway,below, sloped ] {$\alpha$\\[.2em]}  (H);
 
 \draw [->]     (I) -- node [near end,above,sloped ] {$\delta_I I$\\[.2em]}  (D);
 
 \draw [->]     (H) -- node [midway,above,sloped ] {$\delta_H H$\\[.2em]}  (D);
 
 \draw [->]     (H) -- node [near start,above,sloped ] {$\gamma_H H$\\[.2em]}  (R);
 
 \node [draw,text width=6cm] (idr) at (1,-6.5) {\textbf{Identifiability Results:}\\\textbf{Without Initial Conditions}\\ identifiable parameters: None \\ unidentifiable parameters: $\beta_I,\beta_H,\beta_D,k,\alpha,\delta_I,\delta_H,\gamma_H,\gamma_I$ \\
  parameter correlations: See equation \eqref{cp1} and \eqref{cp2} };
   \node [draw,text width=6cm] (idr) at (8,-6.5) {\textbf{Identifiability Results:}\\\textbf{With Initial Conditions}\\ identifiable parameters: DAISY did not give any results for this case. \\ };
   
\end{tikzpicture}
\end{figure}
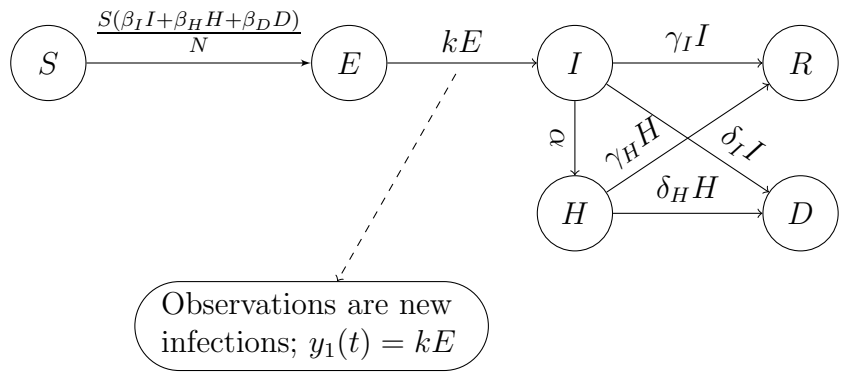

Solving  $c(\bfa p) = c(\bfa {\hat{p}})$ for the model parameters, we obtain the same correlations \eqref{cp1} and \eqref{cp2}. Therefore, we conclude that the model \eqref{Model7} is not structured to reveal all its parameters from the number of newly infected cases alone.
\begin{pro}
The model \ref{Model7} is not structurally identifiable from the observations of newly infected cases.
\end{pro}
We note that DAISY does not yield results when all initial conditions are known.The enhanced compartmental diagram and results of structural identifiability of model \ref{Model7}  when observations are new infections is shown in Figure \ref{fig:model7a}. \\

\underline{\textbf{(ii) Number of new infections and hospitalizations are observed:}}\\
Next, we consider the scenario when data on the number of new hospitalizations are also available. That is, we observe both the number of new infections  $y_1(t) = k E (t)$ and new hospitalizations $y_2(t)=\alpha I(t).$ In this scenario, DAISY yields the following input-output equations:
\begin{equation*}
\begin{aligned}
0&= y_1'- \frac{y_1}{\beta_I }( - \alpha\beta_H + \alpha \beta_I - \beta_D\delta_I - \beta_I\delta_H +
\beta_I\delta_I - \beta_I\gamma_H + \beta_I\gamma_I) -\frac{y_2}{\alpha\beta_I} (\alpha^2 \beta_H - \\&\alpha^2 \beta_I - \alpha \beta_D \delta_H + \alpha
\beta_D\delta_I + \alpha\beta_H\delta_I + \alpha\beta_H\gamma_I + \alpha\beta_I\delta_H - 2\alpha\beta_I\delta_I + \alpha
\beta_I\gamma_H - 2\alpha\beta_I\gamma_I -\\& \beta_D\delta_H\delta_I + \beta_D\delta_I^2 - \beta_D\delta_I\gamma_H + \beta_D
\delta_I\gamma_I + \beta_I\delta_H\delta_I + \beta_I\delta_H\gamma_I - \beta_I\delta_I^2 + \beta_I\delta_I\gamma_H - 2\beta_I
\delta_I\gamma_I +\\& \beta_I\gamma_H\gamma_I - \beta_I\gamma_I^2)\\
0&= y_2' - y_1\alpha + y_2(\alpha + \delta_I + \gamma_I)
\end{aligned}
\end{equation*}
We solve $c(\bfa p) = c(\hat{ \bfa p})$ and obtain
\begin{equation} \label{res_7b}
\begin{aligned}
\{\alpha&=\hat{\alpha},\quad \gamma_I
+\delta_I=+\hat{\delta}_I+\hat{\gamma}_I,\quad \beta_H= \frac{1}{\hat{\alpha} \beta_D \hat{\beta_I} \delta_I}(\hat{\alpha} \beta_D \beta_I \hat{\beta_I} \delta_H -\hat{\alpha} 
\hat{\beta}_D^2 \hat{\delta}_H
+\hat{\alpha} \beta_D \hat{\beta}_H \beta_I \delta_I +\\& 
  \beta_D \beta_I \hat{\beta_I} \hat{\delta}_H \delta_I - \beta_D^2 \hat{\beta_I} \delta_I^2 - \hat{\beta}_D \beta_I^2 \hat{\delta}_H \hat{\delta_I} + 
  \beta_D \hat{\beta}_D \beta_I \delta_I \hat{\delta_I} + 
  \beta_D \beta_I \hat{\beta_I} \delta_I \hat{\gamma}_H - \hat{\beta_D} \beta_I^2 \hat{\delta_I}\hat{\gamma}_H), \quad  \\&\gamma_H=\frac{1}{\beta_D \hat{\beta_I} \delta_I}
  (-\hat{\alpha} \beta_D \hat{\beta_I} \delta_H + \hat{\alpha} \hat{\beta}_D \beta_I \hat{\delta}_H -
   \beta_D \hat{\beta_I} \delta_H \delta_I + \hat{\beta}_D \beta_I \hat{\delta}_H \hat{\delta_I} +
   \hat{\beta}_D \beta_I \hat{\delta_I} \hat{\gamma}_H)\}
\end{aligned}
\end{equation}

\begin{figure}
\caption{Compartmental diagram for Model \ref{Model7}. Circles show the epidemiological compartments for the different states of the system. Solid arrows indicate the transitions between compartments. The dashed arrows indicate the source of the observations, which are the newly infected cases and new hospitalizations. Structural identifiability analysis results are summarized for the scenario when initial conditions are known and unknown.}\label{fig:model7b}
\tikzstyle{line} = [draw, -latex']
\begin{tikzpicture} 
  
  \node [circle,draw,minimum size=1cm] (S) at (0,0) {$S$}; 
  \node [circle,draw,minimum size=1cm]  (E) at (4,0) {$E$}; 
  \node [circle,draw,minimum size=1cm]  (I) at (7,0) {$I$}; 
  \node [circle,draw,minimum size=1cm]  (R) at (10,0) {$R$}; 
   \node [circle,draw,minimum size=1cm]  (H) at (7,-2) {$H$}; 
  \node [circle,draw,minimum size=1cm]  (D) at (10,-2) {$D$};
  
  \node [draw,rounded rectangle,text width=3cm] (m) at (4.5,2) {Observations are new infections; $y_1(t) = kE(t)$};
  
 \node [draw,rounded rectangle,text width=4cm] (O) at (3.5,-3.5) {Observations are new hospitalizations; $y_2(t) = \alpha I$};
 
 \node (z) at (5,0) {};
 \node (zz) at (6.50,-1) {};
 \draw[->, dashed] (zz) --(O);
 \draw[->, dashed] (z) --(m);

 \path [line]    (S) -- node [midway,above ] {$\frac{S(\beta_I I+\beta_H H+\beta_D D)}{N}$}  (E); 
 \draw [->]     (E) -- node [midway,above,sloped ] {$kE$\\[.2em]}  (I);
 
 \draw [->]     (I) -- node [midway,above,sloped ] {$\gamma_I I$\\[.2em]}  (R);
 
 \draw [->]     (I) -- node [midway,below, sloped ] {$\alpha I$\\[.2em]}  (H);
 
 \draw [->]     (I) -- node [near end,above,sloped ] {$\delta_I I$\\[.2em]}  (D);
 
 \draw [->]     (H) -- node [midway,above,sloped ] {$\delta_H H$\\[.2em]}  (D);
 
 \draw [->]     (H) -- node [near start,above,sloped ] {$\gamma_H H$\\[.2em]}  (R);
 
 \node [draw,text width=6cm] (idr) at (1,-6.5) {\textbf{Identifiability Results:}\\\textbf{Without Initial Conditions}\\ identifiable parameters: $\alpha$ \\ unidentifiable parameters:
 $k$,$\beta_I$,$\beta_H$,$\beta_D$,$k$,$\alpha$,$\delta_I$,$\delta_H$,$\gamma_H$,$\gamma_I$ \\
  parameter correlations: see \eqref{res_7b} };
   \node [draw,text width=6cm] (idr) at (8,-6.5) {\textbf{Identifiability Results:}\\\textbf{With Initial Conditions}\\ identifiable parameters: DAISY did not give any results for this case. \\ };
   
\end{tikzpicture}
\end{figure}
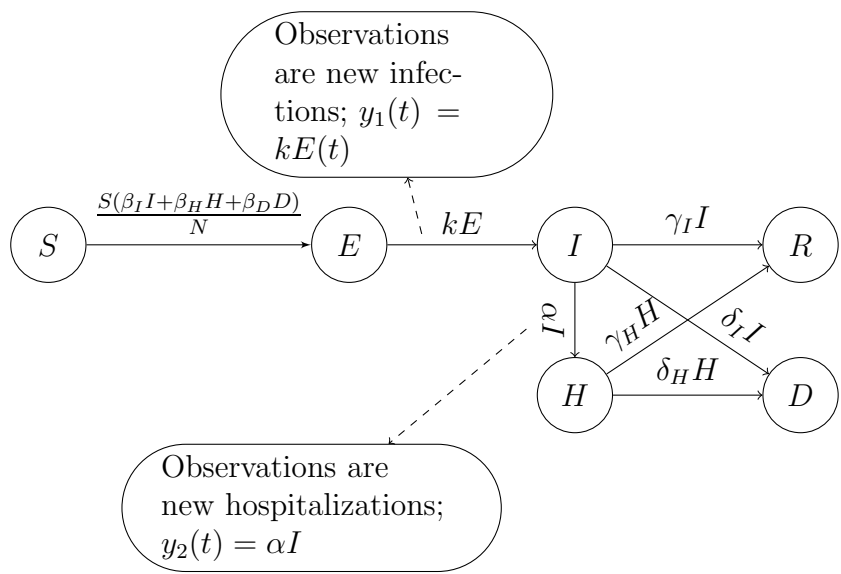

As seen from \eqref{res_7b}, we can only identify the parameter $\alpha$ and the sum $\gamma_I + \delta_I$, while the rest of the parameters are involved in complex correlations.  Furthermore, the parameter $k$ does not appear in the input-output equations, therefore is not identifiable. Hence, we conclude that the model \ref{Model7} is not identifiable from the observations of newly infected cases and hospitalizations. 
\begin{pro}
The model \ref{Model7} is not structurally identifiable from the number of newly infected cases and hospitalizations.
\end{pro}
The enhanced compartmental diagram and results of structural identifiability of model \ref{Model7}  when observations are new infections and hospitalizations is shown in Figure \ref{fig:model7b}. 

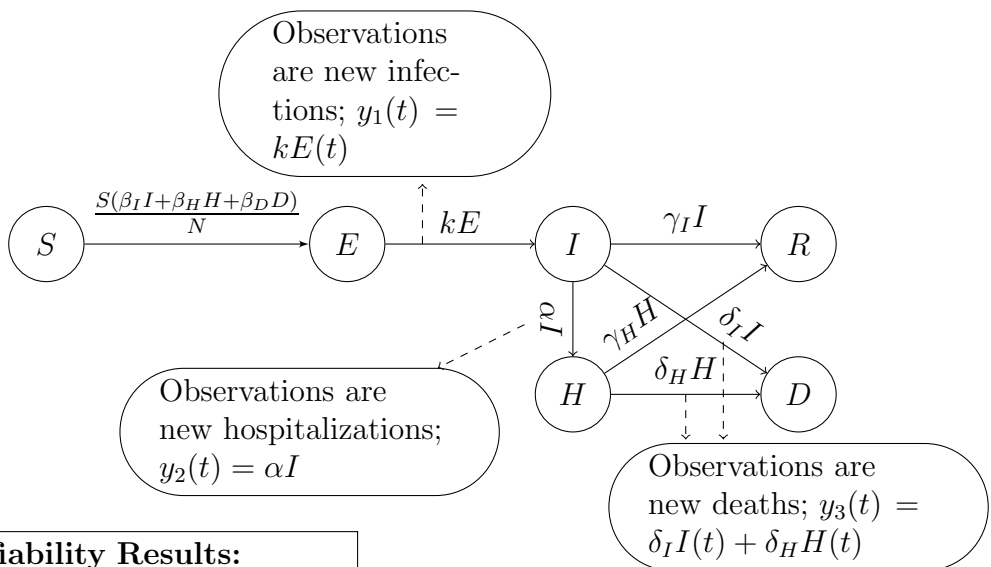
\begin{figure}
\caption{Compartmental diagram for model \ref{Model7}. Circles show the epidemiological compartments for the different states of the system. Solid arrows indicate the transitions between compartments. The dashed arrows indicate the source of the observations, which are the new cases, new hospitalizations and new deaths. Structural identifiability analysis results are summarized for the scenarios when the initial conditions are known and unknown.}\label{fig:model7c}
\tikzstyle{line} = [draw, -latex']
\begin{tikzpicture} 
  
  \node [circle,draw,minimum size=1cm] (S) at (0,0) {$S$}; 
  \node [circle,draw,minimum size=1cm]  (E) at (4,0) {$E$}; 
  \node [circle,draw,minimum size=1cm]  (I) at (7,0) {$I$}; 
  \node [circle,draw,minimum size=1cm]  (R) at (10,0) {$R$}; 
   \node [circle,draw,minimum size=1cm]  (H) at (7,-2) {$H$}; 
  \node [circle,draw,minimum size=1cm]  (D) at (10,-2) {$D$};
  
  \node [draw,rounded rectangle,text width=3cm] (m) at (4.5,2) {Observations are new infections; $y_1(t) = kE(t)$};
  
 \node [draw,rounded rectangle,text width=4cm] (O) at (3.5,-2.5) {Observations are new hospitalizations; $y_2(t) = \alpha I$};
 
 \node [draw,rounded rectangle,text width=4cm] (kk) at (10,-3.5) {Observations are new deaths; $y_3(t)=\delta_I I(t)+\delta_H H(t)$};
 
 \node (z) at (5,0) {};
 \node (zz) at (6.50,-1) {};

 \draw[->, dashed] (zz) --(O);
 \draw[->, dashed] (9,-1.3) --(9,-2.6);
 \draw[->, dashed] (8.5,-2) --(8.5,-2.6);
 \draw[->, dashed] (5,0) --(5,0.8);

 \path [line]    (S) -- node [midway,above ] {$\frac{S(\beta_I I+\beta_H H+\beta_D D)}{N}$}  (E); 
 \draw [->]     (E) -- node [midway,above,sloped ] {$kE$\\[.2em]}  (I);
 
 \draw [->]     (I) -- node [midway,above,sloped ] {$\gamma_I I$\\[.2em]}  (R);
 
 \draw [->]     (I) -- node [midway,below, sloped ] {$\alpha I$\\[.2em]}  (H);
 
 \draw [->]     (I) -- node [near end,above,sloped ] {$\delta_I I$\\[.2em]}  (D);
 
 \draw [->]     (H) -- node [midway,above,sloped ] {$\delta_H H$\\[.2em]}  (D);
 
 \draw [->]     (H) -- node [near start,above,sloped ] {$\gamma_H H$\\[.2em]}  (R);
 
 \node [draw,text width=6cm] (idr) at (1,-6.5) {\textbf{Identifiability Results:}\\\textbf{Without Initial Conditions}\\ identifiable parameters: $\alpha,\delta_I,\gamma_I, \delta_H, \gamma_H$ \\ unidentifiable parameters: $k,\beta_I,\beta_H,\beta_D$ \\
  parameter correlations: $\D\frac{\beta_D}{\beta_H}=\D\frac{\hat{\beta}_D}{\hat{\beta}_H}$, $\D\frac{\beta_I}{\beta_H}=\D\frac{\hat{\beta_I}}{\hat{\beta_H}}.$ };
   \node [draw,text width=6cm] (idr) at (8,-6.5) {\textbf{Identifiability Results:}\\\textbf{With Initial Conditions}\\ identifiable parameters: DAISY did not give any results for this case. \\ };
   
\end{tikzpicture}
\end{figure}

\underline{\textbf{(iii) Number of new infections, hospitalizations and deaths are}}\\ \underline{\textbf{observed:}} \\
We now consider the scenario when the number of new deaths from the disease are available in addition to the number of new cases and hospitalizations. Hence, here we observe three datasets: the number of newly infected  cases $y_1(t)=k E(t)$, the number of new hospitalizations $y_2(t)=\alpha I(t)$, and the number of new deaths $y_3(t)=\delta_I I(t)+\delta_H H(t)$. Using DAISY, we obtain the following input-output equations:
\begin{equation}
\begin{aligned}
0&=y_2'- y_1\alpha + y_2(\alpha + \delta_I + \gamma_I)\\
0&=y_3'- y_1\delta_I+ \frac{y_2}{\alpha\beta_H }(\alpha\beta_H\delta_I- \beta_D\delta_H\delta_I- \beta_H
\delta_H\delta_I+ \beta_H\delta_I^2 + \beta_H\delta_I\gamma_I + \beta_I\delta_H^2) +\\ 
&y_3\frac{\beta_D\delta_H}{\beta_H}\\
0&= y_1'- \frac{y_3''}{\delta_I }+ y_1(\alpha - \delta_H +
\delta_I + \gamma_I) - \frac{y_2}{\alpha\beta_H\delta_I} ( - \alpha^2 \beta_H \delta_I - \alpha \beta_H \delta_H^2 + \alpha \beta_H \delta_H \delta_I -\\
&2\alpha
\beta_H\delta_I^2
 - 2\alpha \beta_H\delta_I\gamma_I + \beta_D \delta_H^2 \delta_I + \beta_D \delta_H \delta_I \gamma_H + \beta_H \delta_H \delta_I^2 + \beta_H \delta_H \delta_I \gamma_I - \beta_H \delta_I^3 -\\ 
&2\beta_H \delta_I^2 \gamma_I -
  \beta_H \delta_I
\gamma_I^2 - \beta_I \delta_H^3 - \beta_I \delta_H^2\gamma_H) + y_3\frac{\beta_D \delta_H }{\beta_H\delta_I}(\delta_H + \gamma_H)
\end{aligned}
\end{equation}
As before, we first set  $c(\bfa p) = c(\hat{\bfa p})$, and solve for the model parameters and obtain the following parameter correlations,
$$\{\alpha=\hat{\alpha},\quad \delta_I=\hat{\delta}_I,\quad \gamma_I=\hat{\gamma}_I,\quad \delta_H=\hat{\delta}_H,\quad
\gamma_H=\hat{\gamma}_H, \quad \frac{\beta_D}{\beta_H}=\frac{\hat{\beta}_D}{\hat{\beta}_H},\quad\frac{\beta_I}{\beta_H}=\frac{\hat{\beta}_I}{\hat{\beta}_H}\}$$
We can only identify parameters $\alpha,\delta_I,\gamma_I, \delta_H, \gamma_H$ and the ratios $\D\frac{\beta_D}{\beta_H}$, $\D\frac{\beta_I}{\beta_H}.$ The parameter $k$ is not identifiable as it does not appear in the input-output equations. This implies that the model \ref{Model7} is not structurally identifiable given the number of new cases, new hospitalizations and new deaths observations. 
\begin{pro}
The model \ref{Model7} is not structurally identifiable from the number of newly infected cases, new hospitalizations, and new deaths.
\end{pro}
The enhanced compartmental diagram and results of structural identifiability of model \ref{Model7}  when observations are new infections, hospitalizations and deaths is shown in 
Figure \ref{fig:model7c}. \\

If we assume that transmission stemming from the death compartment is negligible, we can eliminate the term $\beta_D D(t)S(t)/N$ and obtain a simpler version of the model \eqref{Model7}:
\begin{equation}\label{Model7'}\tag{M$_7'$}
\textbf{Model 7' (M$_7'$):}
\begin{cases}
\displaystyle\frac{dS}{dt}= \displaystyle\frac{-(\beta_{I} I + \beta_{H} H) S}{N}, \quad S(0) = S_0\\[1.5ex]
 \displaystyle\frac{dE}{dt}=  \displaystyle\frac{(\beta_{I} I + \beta_{H} H) S}{N}-kE, \quad E(0) = E_0\\[1.5ex]
\displaystyle\frac{dI}{dt}=kE-(\alpha+\gamma_{I}+\delta_{I})I, \quad I(0) = I_0\\[1.5ex]
\displaystyle\frac{dH}{dt}=  \alpha I -(\gamma_{H}+\delta_{H}) H, \quad H(0) = H_0\\[1.5ex]  
\displaystyle\frac{dR}{dt}=  \gamma_{I} I+\gamma_{H} H, \quad R(0) = R_0\\[1.5ex]
 \displaystyle\frac{dD}{dt}=  \delta_{I} I+\delta_{H} H, \quad D(0) = D_0.\\[1.5ex]
 \end{cases}
\end{equation}
Using DAISY, we obtain the following result,
$$\{\beta_I=\hat{\beta_I},\quad \beta_H=\hat{\beta}_H,\quad k=\hat{k},\quad \alpha=\hat{\alpha},\quad \gamma_I=\hat{\gamma_I},\quad\delta_I=\hat{\delta_I}\quad\gamma_H=\hat{\gamma}_H,\quad \delta_H=\hat{\delta}_H\}$$

Hence, the modified model becomes structurally identifiable and reach the following conclusion.

\begin{pro}
The model \ref{Model7'} is structurally identifiable from the number of newly infected cases, new hospitalizations, and new deaths.
\end{pro}
The enhanced compartmental diagram and results of structural identifiability of model \ref{Model7'}  when observations are new infections, hospitalizations and deaths is shown in Figure \ref{fig:model7d}. 

\begin{figure}
\caption{Compartmental diagram for Model \ref{Model7'}. Circles show the epidemiological compartments for the different states of the system. Solid arrows indicate the transitions between compartments. The dashed arrows indicate the source of the observations, which are the cases, new hospitalizations and new deaths. Structural identifiability analysis results are summarized for the scenario when the initial conditions are unknown.}\label{fig:model7d}
\tikzstyle{line} = [draw, -latex']
\begin{tikzpicture} 
  
  \node [circle,draw,minimum size=1cm] (S) at (0,0) {$S$}; 
  \node [circle,draw,minimum size=1cm]  (E) at (4,0) {$E$}; 
  \node [circle,draw,minimum size=1cm]  (I) at (7,0) {$I$}; 
  \node [circle,draw,minimum size=1cm]  (R) at (10,0) {$R$}; 
   \node [circle,draw,minimum size=1cm]  (H) at (7,-2) {$H$}; 
  \node [circle,draw,minimum size=1cm]  (D) at (10,-2) {$D$};
  
  \node [draw,rounded rectangle,text width=3cm] (m) at (4.5,2) {Observations are new infections; $y_1(t) = kE(t)$};
  
 \node [draw,rounded rectangle,text width=4cm] (O) at (3.5,-2.5) {Observations are new hospitalizations ; $y_2(t) = \alpha I$};
 
 \node [draw,rounded rectangle,text width=4cm] (kk) at (10,-3.5) {Observations are new deaths.; $y_3(t)=\delta_I I(t)+\delta_H H(t)$};
 
 \node (z) at (5,0) {};
 \node (zz) at (6.50,-1) {};
 \draw[->, dashed] (zz) --(O);
 \draw[->, dashed] (9,-1.3) --(9,-2.6);
 \draw[->, dashed] (8.5,-2) --(8.5,-2.6);
 \draw[->, dashed] (5,0) --(5,0.8);

 \path [line]    (S) -- node [midway,above ] {$\frac{S(\beta_I I+\beta_H H)}{N}$}  (E); 
 \draw [->]     (E) -- node [midway,above,sloped ] {$kE$\\[.2em]}  (I);
 
 \draw [->]     (I) -- node [midway,above,sloped ] {$\gamma_I I$\\[.2em]}  (R);
 
 \draw [->]     (I) -- node [midway,below, sloped ] {$\alpha I$\\[.2em]}  (H);
 
 \draw [->]     (I) -- node [near end,above,sloped ] {$\delta_I I$\\[.2em]}  (D);
 
 \draw [->]     (H) -- node [midway,above,sloped ] {$\delta_H H$\\[.2em]}  (D);
 
 \draw [->]     (H) -- node [near start,above,sloped ] {$\gamma_H H$\\[.2em]}  (R);
 
 \node [draw,text width=6cm] (idr) at (1,-6.5) {\textbf{Identifiability Results:}\\\textbf{Without Initial Conditions}\\ identifiable parameters: $\alpha,\delta_I,\gamma_I, \delta_H, \gamma_H, k,\beta_I,\beta_H,\beta_D$ \\ unidentifiable parameters: None \\
  parameter correlations: None };
   
\end{tikzpicture}
\end{figure}

\subsection{COVID-19 model (\ref{Model8})}

This model is motivated by the transmission dynamics of some respiratory infections such as COVID-19, where pre-symptomatic and symptomatic cases contribute to the transmission process \citep{TuncerCovid}. The model is given by the following system of differential equations:

\begin{equation}\label{Model8}\tag{M$_8$}
\textbf{Model 8 (M$_8$):}
\begin{cases}
\displaystyle\frac{dS}{dt} = -\frac{(\beta_{\rho} I_{\rho} + \beta_{I} I) S}{N}, \quad S(0) = S_0\\[1.5ex]
\displaystyle\frac{dE}{dt} = \frac{(\beta_{\rho} I_{\rho} + \beta_{I} I) S}{N}-kE, \quad E(0) = E_0\\[1.5ex]
\displaystyle\frac{dI_{\rho}}{dt}= kE - k_\rho I_{\rho} -  \gamma_{\rho} I_{\rho}, \quad I_{\rho}(0) = I_{\rho 0}\\[1.5ex]
\displaystyle\frac{dI}{dt} = k_\rho  I_{\rho}-\gamma I - \delta I, \quad I(0) = I_0\\[1.5ex]
\displaystyle\frac{dR}{dt} =  \gamma I +\gamma_{\rho} I_\rho, \quad R(0) = R_0\\[1.5ex]
\displaystyle\frac{dD}{dt} = \delta I, \quad D(0) = D_0.\\[1.5ex]
\end{cases}
\end{equation}

This epidemic model involves six epidemiological states: The number of susceptible individuals $(S(t))$, the number of latent individuals $(E(t))$, the number of pre-symptomatic individuals $(I_\rho(t))$, the number of symptomatic individuals $(I(t))$, the number of recovered individuals $(R(t))$ and the number of deaths $(D(t)).$ Unlike the previous models, this model has a pre-symptomatic infectious state. Susceptible individuals get infected by both pre-symptomatic and symptomatic individuals at the rate $\beta_\rho I_\rho(t)/N + \beta_I I(t)/N$. The set of model parameters is given by $\bfa p=(\beta_\rho,\beta_I, k, k_\rho,\gamma, \gamma_\rho, \delta)$. We will investigate the structural identifiability of the model \ref{Model8} in two different observation settings; (i) number of new symptomatic cases is observed, (ii) number of new symptomatic cases  and deaths  are observed.

\underline{\textbf{(i) Number of new symptomatic cases is observed}}. We first consider the scenario where the observations correspond to the number of new symptomatic cases,  $y_1(t) = k_{\rho} I_{\rho}(t)$. In this scenario, DAISY cannot derive a result (DAISY input code available on our GitHub site). It was computationally very expensive for DAISY to derive the characteristic set of  model \ref{Model8} and observations, $y_1(t)$.  Therefore,  we  consider additional observations in the stuctural identifiability analysis. 

\underline{\textbf{(ii) Number of new symptomatic cases  and deaths  are observed}}. Next, we analyze the scenario when the observations correspond to both the number of new cases and deaths:
$$y_1(t) = k_{\rho}  I_{\rho}(t) \quad \text{and}\quad y_2(t) = \delta I(t).$$ This time, DAISY derives the input-output equations (DAISY result file is available from our GitHub site). DAISY assigns the following integers to the model parameters,
$$\hat{\bfa p} = \left\{\hat{\beta}_\rho = 2,\; \hat{\beta_I} =3, \; \hat{k}=5,\; \hat{k_\rho}=7,\; \hat{\gamma}_\rho = 11,\hat{\gamma} = 13,\hat{\delta} = 17\right\}$$ and then solves 
$c(\bfa p) = c(\hat{\bfa p})$ and gives the following results:
\begin{equation}\label{D_res_m8}
7\beta_\rho + 2\gamma_\rho= 36,
\quad \delta=17, \quad \gamma=13, \quad k=5, \quad \beta_I=3,
\quad k_\rho+\gamma_\rho=18.
\end{equation}
and
\begin{equation}\label{D_res_m8_2}
7\beta_\rho + 2\gamma_\rho= 10, 
\quad \delta=17, \quad \gamma=13, \quad k=18, \quad \beta_I=3,
\quad k_\rho+\gamma_\rho=5.
\end{equation}
In this case, DAISY yields two possible solution sets given in \eqref{D_res_m8} and \eqref{D_res_m8_2}. While the parameters $\gamma$, $\delta$, and $\beta_I$ are identifiable, parameters $\beta_\rho$, $\gamma_\rho$ and $k_\rho$  are correlated and  not identifiable. Parameter $k$  may have 2 different solutions, $k= \hat{k}$ which is the true solution or $k = \hat{k}_\rho + \hat{\gamma}_\rho$. Even in the case when $k$ is identified correctly, which is the solution set \eqref{D_res_m8}, several correlations involve parameters $\beta_\rho, \gamma_\rho$  and $k_\rho $.
Therefore, we reach the following proposition for the structural identifiability of \ref{Model8} from the observations of the number of new symptomatic cases and deaths. 

\begin{pro} \label{prop1_m8}
The model \ref{Model8} is not structurally identifiable from the observation of new symptomatic cases and  deaths. Only the parameters $\gamma$, $\delta$, $\beta_I$ are identifiable. 
\end{pro}

To obtain a structurally identifiable model, we need to first fix $k=\hat{k}$, this would restrict the solution to the set \eqref{D_res_m8}. We then set $\gamma_\rho = \hat{\gamma}_\rho$ in order to diminish the correlations between the parameters $\beta_\rho$, $\gamma_\rho$ and $k_\rho$. Alternatively, fixing $\beta_\rho$, or  $k_\rho$
would also break down the correlations.
We summarize the result in the  Proposition \ref{prop1_m8_2}. 

\begin{pro} \label{prop1_m8_2}
If the pre-symptomatic recovery rate, $\gamma_\rho$, and the latent period, $1/k$, are fixed, then the model \ref{Model8} becomes structurally identifiable from the observation of new symptomatic cases  and deaths. 
\end{pro}

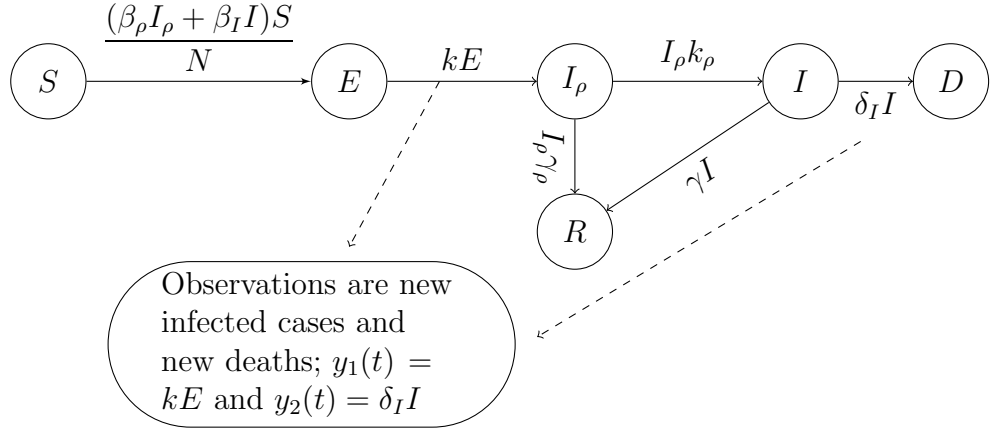
\begin{figure}
\caption{Compartmental diagram for Model \ref{Model8}. Circles show the epidemiological compartments for the different states of the system. Solid arrows indicate the transitions between compartments. The dashed arrows indicate the source of the observations, which are the new infected cases and new deaths. Structural identifiability analysis results are summarized when initial conditions are known and unknown.}\label{fig:model8}
\tikzstyle{line} = [draw, -latex']
\begin{tikzpicture} 
  
  \node [circle,draw,minimum size=1cm] (S) at (0,0) {$S$}; 
  \node [circle,draw,minimum size=1cm]  (E) at (4,0) {$E$}; 
  \node [circle,draw,minimum size=1cm]  (I_p) at (7,0) {$I_\rho$}; 
    \node [circle,draw,minimum size=1cm]  (I) at (10,0) {$I$}; 
   \node [circle,draw,minimum size=1cm]  (R) at (7,-2) {$R$}; 
  \node [circle,draw,minimum size=1cm]  (D) at (12,0) {$D$};
  
  \node [draw,rounded rectangle,text width=4cm] (O) at (3.5,-3.5) {Observations are new infected cases and new deaths; $y_1(t) = kE$ and $y_2(t) = \delta_I I$};

  \draw[->, dashed] (5.2,0) --(4,-2.2);
 \draw[->, dashed] (10.8,-0.8) --(6.5,-3.4);

 \path [line]    (S) -- node [midway,above ] {$\D\frac{(\beta_\rho I_\rho +\beta_I I)S}{N}$}  (E); 
 
 \draw [->]     (E) -- node [midway,above,sloped ] {$k E$\\[.2em]}  (I_p);
 
 \draw [->]     (I_p) -- node [midway,above,sloped ] {$I_\rho k_\rho $\\[.2em]}  (I);
 
 \draw [->]     (I_p) -- node [midway,below,sloped ] {$I_\rho \gamma_\rho $\\[.2em]}  (R);
 
 \draw [->]     (I) -- node [midway,below,sloped ] {$\gamma I $\\[.2em]}  (R);

 \draw [->]     (I) -- node [midway,below, sloped ] {$\delta_I I$\\[.2em]}  (D);
 
 \node [draw,text width=6cm] (idr) at (1,-7.5) {\textbf{Identifiability Results:}\\\textbf{Without Initial Conditions}\\ identifiable parameters: $\gamma$, $\delta$, $\beta_I$  \\ unidentifiable parameters: $\beta_\rho$, $\gamma_\rho$, $k$ and $k_\rho$ \\
  parameter correlations: Two sets of correlation shown in \eqref{D_res_m8} and \eqref{D_res_m8_2} };
   \node [draw,text width=6cm] (idr) at (8,-6.5) {\textbf{Identifiability Results:}\\\textbf{With Initial Conditions}\\ identifiable parameters: $\gamma$, $\delta$, $\gamma_\rho$, $k$, $k_\rho$, $\beta_\rho$ and $\beta_I$ \\ };
   
\end{tikzpicture}
\end{figure}

Next, we carry out the structural identifiability analysis of the model \ref{Model8} from the observation of new cases and deaths when all the initial conditions are known. With known initial conditions, DAISY solves $c(\bfa p) = c(\hat{\bfa p})$ together with the characteristic set evaluated at time $t=0$ and 
provides the following solution  
$\gamma=13$, $\beta_\rho =2$, $\gamma_\rho=11$, $\beta_I=3$, $\delta=17$,
$k=5$, and $k_\rho=7$.
Therefore, we conclude that the model is structurally identifiable when the initial conditions are known and reach the following proposition.
\begin{pro}
The model \ref{Model8} is structurally identifiable from the observation of new cases and deaths when all the initial conditions are known.
\end{pro}
The enhanced compartmental diagram of model \ref{Model8} that includes the observed states of the system and the parameter correlations resulting from the structural identifiability analysis is given in Figure \ref{fig:model8}. 

\begin{table}[h!]
    \begin{center}
    \footnotesize
    \hspace*{-2cm}
    \begin{tabular}{l|l|l}
    \textbf{Notation}&\textbf{Meaning}&\textbf{Used in model}\\
        \hline
        $S(t)$ & Number of susceptible individuals at time $t$ & \ref{Model1}, \ref{Model2}, \ref{Model3}, \ref{Model4}, \ref{Model5}, \ref{Model6}, \ref{Model7}, \ref{Model8} \\ \hline
        $S_\upsilon(t)$ & Number of susceptible mosquitoes at time $t$ & \ref{Model5}, \ref{Model6} \\ \hline
        $E(t)$ & Number of exposed individuals at time t & \ref{Model1}, \ref{Model2}, \ref{Model3}, \ref{Model4}, \ref{Model7}, \ref{Model8} \\ \hline
        $I(t)$ & Number of infected individuals at time t & \ref{Model1}, \ref{Model2}, \ref{Model3}, \ref{Model4}, \ref{Model5}, \ref{Model6}, \ref{Model7},\ref{Model8} \\ \hline
        $I_\upsilon(t)$ & Number of infected mosquitoes at time t & \ref{Model5}, \ref{Model6}\\ \hline
        $I_\rho(t)$ & Number of pre-symptomatic individuals at time t & \ref{Model8}\\ \hline
       $A(t)$ & Number of asymptomatic individuals at time t & \ref{Model2}, \ref{Model3}, \ref{Model4}, \ref{Model6} \\ \hline
        $R(t)$ & Number of recovered individuals at time t & \ref{Model1}, \ref{Model2}, \ref{Model3}, \ref{Model5}, \ref{Model6}, \ref{Model7}, \ref{Model8} \\ \hline
        $H(t)$ & Number of hospitalized individuals at time t & \ref{Model7} \\ \hline
        $D(t)$ & Number of dead individuals at time t & \ref{Model4}, \ref{Model7}, \ref{Model8} \\ \hline
        
        \end{tabular}
    \end{center}
    \caption{State variables and their definitions for Models  \ref{Model1} through \ref{Model8} covered in the tutorial.}
    \label{Table 1}
\end{table}

\begin{table}[h!]
    \begin{center}
    \footnotesize
    \hspace*{-2cm}
    \begin{tabular}{l|l|l}
    \textbf{Notation}&\textbf{Meaning}&\textbf{Used in model}\\
        \hline
       $\beta$ & Transmission rate in human population & \ref{Model1}, \ref{Model2}, \ref{Model4}, \ref{Model5} \\ \hline
        $\beta_I$ & Transmission rate in infected human population & \ref{Model3}, \ref{Model6}, \ref{Model7}, \ref{Model8}  \\ \hline
        $\beta_\rho$ & Transmission rate in pre-symptomatic human population & \ref{Model8}  \\ \hline
        $\beta_A$ & Transmission rate in asymptomatic individuals & \ref{Model3},  \ref{Model6}, \ref{Model8}  \\ \hline
        $\beta_\upsilon$ & Transmission rate in mosquito & \ref{Model5} \\ \hline
        $\beta_H$ & Transmission rate in hospitalized individuals & \ref{Model7}  \\ \hline
        $\beta_D$ & Transmission rate from dead individuals & \ref{Model7}  \\ \hline
        k & Latent period & \ref{Model1}, \ref{Model2}, \ref{Model3}, \ref{Model4}, \ref{Model7}, \ref{Model8}  \\ \hline
        $k_\rho$ & The fraction of pre-symptomatic individuals who develop symptom & \ref{Model8}  \\ \hline
        $\rho$ & The fraction of infected individuals who show symptoms (always less than 1) & \ref{Model2}, \ref{Model3}, \ref{Model4}, \ref{Model6}\\ \hline
        $\gamma$ & Recovery rate & \ref{Model1}, \ref{Model2}, \ref{Model3}, \ref{Model4}, \ref{Model5}, \ref{Model6}\\ \hline
        $\gamma_I$ & Recovery rate for infected& \ref{Model7}\\ \hline
        $\gamma_H$ & Recovery rate for hospitalized& \ref{Model7}\\ \hline
        $\gamma_\rho$ & The fraction of pre-symptomatic individuals who recover & \ref{Model8}\\ \hline
        $\delta$ & Death rate due to disease among individuals & \ref{Model4}, \ref{Model7}\\ \hline
        $\delta_H$ & Death rate among hospitalized individuals & \ref{Model7}\\ \hline
        $\delta_I$ & Death rate among infected individuals & \ref{Model7}\\ \hline
        $\Lambda_\upsilon$ & Rate at which mosquitoes enter the susceptible class & \ref{Model5}, \ref{Model6}\\ \hline
        $\mu_\upsilon$ & Natural death rate of mosquitoes &  \ref{Model5}, \ref{Model6}\\ \hline
        $\alpha$ & Rate of hospitalization of infected individuals & \ref{Model7}\\ \hline
        \end{tabular}
    \end{center}
    \caption{Parameters and their definitions for Models  \ref{Model1} through \ref{Model8} covered in the tutorial.}
    \label{Table 2}
\end{table}

\begin{table}[h!]
	\begin{center}
		\footnotesize
		\hspace*{-2cm}
		\tiny
		\begin{tabular}{l|l|l|l}
			\textbf{Model}&\textbf{Observed states}&\textbf{Identifiable parameters}&\textbf{Unidentifiable parameters}\\
			\hline
			\ref{Model1} & Number of new infected cases $(kE)$ & $k$, $\gamma$ & $\beta$, $N$ \\ \hline
			\ref{Model2} & Number of new infected symptomatic cases $(k\rho E)$ & $k$, $\gamma$ & $\rho$, $\beta$, $N$ \\ \hline
			\ref{Model3} & Number of new infected symptomatic cases $(k\rho E$) & $k$, $\gamma$ & $\rho$, $\beta_A$, $\beta_I$, $N$ \\ \hline
			\ref{Model4} & Number of new infected cases $(kE)$ & $k$ & $\beta,\delta, \gamma$ \\ \hline
			\ref{Model4} & Number of new infected cases $(kE)$ and new deaths $(\delta I)$ & $k$, $\beta$, $\delta$, $\gamma$  & - \\ \hline
			\ref{Model5} & Cumulative number of new incidence $(S_0-S_t)$ & $\gamma$, $\mu_\upsilon$ & $\beta$, $\beta_\upsilon$, $\Lambda_\upsilon$, $N$ \\ \hline
			\ref{Model6} & Cumulative number of new incidence $\rho$ $(S_0-S_t)$ & $\gamma$,$\mu_\upsilon$,$\rho$ & $\beta_A$, $\beta_I$, $\beta_\upsilon$, $\Lambda_\upsilon$, $N$  \\ \hline
			\ref{Model7} & Number of new infected cases $(kE)$ & - & $\beta_I,\beta_H,\beta_D,k,\alpha,\delta_I,\delta_H,\gamma_H,\gamma_I$ \\ \hline
			\ref{Model7} & \shortstack{Number of new infected cases $(kE)$ and new \\hospitalization $(\alpha I)$} & $\alpha$ & $k,\beta_I,\beta_H,\beta_D,k,\alpha,\delta_I,\delta_H,\gamma_H,\gamma_I$ \\ \hline
			\ref{Model7} & \shortstack{Number of new infected cases $(kE)$, new hospitalization \\$(\alpha I)$ and, new deaths ($\delta_I I+\delta_H H)$}  & $\alpha,\delta_I,\gamma_I, \delta_H, \gamma_H$ & $k,\beta_I,\beta_H,\beta_D$ \\ \hline
			
			\ref{Model8} & Number of new infected cases $(kE)$  & No result obtained from DAISY & $\gamma, \delta, \gamma_\rho, k, k_\rho, \beta_\rho, \beta_I$ \\ \hline
			
			\ref{Model8} & Number of new infected cases $(kE)$ and new deaths ($\delta_I I$) & $\gamma, \delta, k, \beta_I$ & $\beta_\rho, \gamma_\rho, k_\rho$ \\ \hline
			
		\end{tabular}
	\end{center}
	\caption{Summary of the structural identifiability results of the models  \ref{Model1} through \ref{Model8} for different observed states and when initial conditions are not known.}
	\label{Table 3}
\end{table}

\begin{table}[h!] 
	\begin{center}
		\footnotesize
		\hspace*{-2cm}
		\begin{tabular}{l|l|l} 
			\textbf{Model}&\textbf{Observed states}&\textbf{Identifiability result}\\
			\hline
			\ref{Model1} & Number of new infected cases $(kE)$ & All parameters identifiable ($k$, $\gamma$, $\beta$, $N$) \\ \hline
			\ref{Model2} & Number of new infected symptomatic cases $(k\rho E)$ & All parameters identifiable (k, $\gamma$, $\rho, \beta$, $N$)  \\ \hline
			\ref{Model3} & Number of new infected symptomatic cases $(k\rho E)$ & \shortstack{All parameters identifiable ($k$, $\gamma$, $\rho, \beta_A, \beta_I$, $N$)} \\ \hline
			\ref{Model4} & Number of new infected cases $(kE)$ & All parameters identifiable ($k$, $\beta, \delta, \gamma$) \\ \hline
			\ref{Model4} & Number of new infected cases $(kE)$ and new deaths ($\delta I$) & All parameters identifiable ($k$, $\beta, \delta, \gamma$) \\ \hline
			\ref{Model5} & Cumulative number of new incidence $(S_0-S_t)$ & \shortstack{All parameters identifiable ($\gamma$, $\mu_\upsilon$, $\beta$, $\beta_\upsilon$, $\Lambda_\upsilon$, $N$)} \\ \hline
			\ref{Model6} & Cumulative number of new incidence $\rho$ $(S_0-S_t)$ & \shortstack{All parameters identifiable\\ ($\gamma$, $\rho$, $\mu_\upsilon$, $\beta_A$, $\beta_I$, $\beta_\upsilon$, $\Lambda_\upsilon$, $N$)} \\ \hline
			\ref{Model7} & Number of new infected cases ($kE$) & No result obtained from DAISY\\ \hline
			\ref{Model7} & \shortstack{Number of new infected cases $(kE)$ and new \\hospitalization $(\alpha I)$} & No result obtained from DAISY\\ \hline
			\ref{Model7} & \shortstack{Number of new infected cases $(kE)$, new hospitalization \\$(\alpha I)$ and, new deaths $(\delta_I I+\delta_H H)$ }  & No result obtained from DAISY \\ \hline
			\ref{Model8} & Number of new infected cases $(kE)$ & No result obtained from DAISY \\ \hline
			\ref{Model8} & Number of new infected cases $(kE)$ and new deaths ($\delta_I I$) & \shortstack{All parameters identifiable ($\gamma$, $\delta$, k, $\beta_I$, $\beta_\rho$, $\gamma_\rho$, $k_\rho)$} \\ \hline
			
		\end{tabular}
	\end{center}
	\caption{Summary of the structural identifiability results of the models  \ref{Model1} through \ref{Model8} for different observed states and when initial conditions are known.}
	\label{Table 4}
\end{table}	

 \section{Discussion}

In this tutorial paper, we have guided the differential algebra approach using DAISY to conduct structural identifiability analyses of epidemic models based on systems of differential equations when a limited set of state variables is observed. The power of this method lies in its ability to uncover parameter correlations that make a model structurally unidentifiable. We demonstrate how uncovering parameter correlations via structural identifiability analysis can help resolve parameter identifiability roadblocks through various examples. In particular, a structurally identifiable model can be achieved by using the information on the model's initial conditions, revising the structure of the model, obtaining additional information on specific parameters involved in the parameter correlations, or collecting additional data on certain system states. Our results underscore the need to ensure that a free parameter of interest is structurally identifiable before this can be estimated from data. Indeed, achieving a good fit for the data does not imply that the model parameters can be reliably estimated from data. \\

Our results indicate that models' lack of structural identifiability cannot necessarily be remedied by blindly fixing parameters based on existing literature or collecting more data about other system states. It is important to gain insight into how the structure of the model contributes to the lack of parameter identifiability. While several methods exist to conduct structural identifiability analyses of dynamical systems \citep{miao2011identifiability, pohjanpalo1978system, eisenberg2013identifiability, ljung1991testing, chis2011structural, villaverde2016structural, hong2019sian}, the differential algebra approach employed here goes a step further by providing information about existing parameter correlations, which can guide the researcher to resolve parameter identifiability issues. \\

We have also shown how results of structural identifiability analyses help enrich compartmental diagrams of differential-equation models by incorporating two pieces of information: The observed variables (e.g., new cases or new deaths) and the corresponding results of the structural identifiability analysis (i.e., whether parameters are identifiable when the initial conditions are known or not). Importantly, these enhanced compartmental diagrams can feature different parameter identifiability configurations for the same model for different observed state variables. \\

The analysis of structural identifiability using DAISY is not exempt from limitations. First, DAISY uses integers to evaluate the characteristic set numerically, although epidemic model parameters are typically not integers. In some cases, we assessed the characteristic set more generally using \textit{Mathematica}. However, manually transferring long characteristic sets to \textit{Mathematica} is prone to errors. Second, DAISY may incorrectly declare a model structurally not identifiable when the initial conditions are known because DAISY does not know that some of the parameters could be functions of the initial conditions. For this reason, the modeler needs to examine the parameter correlations derived by DAISY to fully evaluate the role of the initial conditions on the structural identifiability of the model parameters. Finally, DAISY may run into computational or memory issues for complex models associated with long characteristic sets and may be unable to produce an output. For situations where the differential algebra approach cannot reveal the structural identifiability properties of the model, alternative approaches could be explored (e.g., \citep{hong2019sian, Gerardo_id}), but these often do not provide explicit answers about existing parameter correlations involved in the dynamical system. \\ 

In summary, this tutorial-based primer will broadly apply to analyze the structural identifiability of epidemic models based on differential equations. The tutorial can be part of the curriculum of student training in mathematical biology, applied statistics, infectious disease modeling, and specialty courses in epidemic modeling. Our analyses indicate that parameter identifiability issues are more likely to arise as model complexity increases in the number of equations/states and parameters. However, incorporating initial conditions and additional observations about other system states may resolve the lack of structural identifiability.

\section{Statements and Declarations}

The authors report no conflicts of interests related to this article. 

 \newpage

 \section{Appendix}
 \newpage
 \subsection{Input-output equations of epidemic models}
 \newpage
 \underline{Input-output equation of the model \ref{Model2}}
 \begin{equation}\label{Model2_input_output}
	\begin{aligned}
		&(y_1^{'''})^2 y_1^{'} y_1 + (y_1^{'''})^2 y_1^2 k - y_1^{'''} (y_1^{''})^2 y_1 - y_1^{'''} y_1^{''} (y_1^{'})^2 +y_1^{'''} y_1^{''} y_1^{'} y_1 (2  \gamma - k) + \\[2.5ex]
		&2 y_1^{'''} y_1^{''} y_1^2 k ( \gamma + k) - y_1^{'''} (y_1^{'})^3 ( \gamma + k) + 4 y_1^{'''} (y_1^{'})^2 y_1^2  \frac{\beta}{N} + y_1^{'''} (y_1^{'})^2 y_1 ( \gamma^2 - 2 k^2) + \\[2.5ex]
		&8 y_1^{'''} y_1^{'} y_1^3  \frac{\beta k}{N} + y_1^{'''} y_1^{'} y_1^2  \gamma k (2  \gamma + k) + 4 y_1^{'''} y_1^4  \frac{\beta k^2}{N} + y_1^{'''} y_1^3  \gamma^2 k^2 + (y_1^{''})^3 y_1^{'}  - (y_1^{''})^3 y_1 ( \gamma + k) +\\[2.5ex]
		&  2 (y_1^{''})^2 (y_1^{'})^2 k - 3 (y_1^{''})^2 y_1^{'} y_1^2  \frac{\beta}{N} + (y_1^{''})^2 y_1^{'} y_1 ( \gamma^2 -  \gamma k - 2 k^2) - 3 (y_1^{''})^2 y_1^3 \frac{\beta k}{N} + \\[2.5ex]
		&  (y_1^{''})^2 y_1^2 k ( \gamma^2 + 2 \gamma k + k^2) + y_1^{''} (y_1^{'})^3 ( - 2 \gamma^2 - \gamma k + 2 k^2) + 2 y_1^{''} (y_1^{'})^2 y_1^2 \frac{\beta (3 \gamma - k)}{N} \\[2.5ex]
		& + y_1^{''} (y_1^{'})^2 y_1 \frac{( \gamma^3 - \gamma^2 k - 3 \gamma k^2 - 2 k^3)}{N^2} + 2 y_1^{''} y_1^{'} y_1^3 \frac{\beta k (6 \gamma + k)}{N} \\[2.5ex]
		& + y_1^{''} y_1^{'} y_1^2 \frac{\gamma k (2 \gamma^2 + 2 \gamma k + k^2)}{N^2}+ 2 y_1^{''} y_1^4 \frac{\beta k^2 (3 \gamma + 2 k)}{N} \\[2.5ex]
		& + y_1^{''} y_1^3 \gamma^2 k^2 ( \gamma + k) - (y_1^{'})^5 \frac{\beta}{N} + (y_1^{'})^4 y_1 \frac{\beta ( - 4 \gamma - 3 k)}{N} \\[2.5ex]
		&+ (y_1^{'})^4 ( - \gamma^3 - 2 \gamma^2 k + k^3) + 4 (y_1^{'})^3 y_1^3 \frac{\beta^2}{N^2} + (y_1^{'})^3 y_1^2 \frac{\beta ( \gamma^2 - 6 \gamma k - 6 k^2)}{N} + \\[2.5ex]
		&(y_1^{'})^3 y_1 \gamma k ( - 2 \gamma^2 - 3 \gamma k - k^2) + 12 (y_1^{'})^2 y_1^4 \frac{\beta^2 k}{N^2} + (y_1^{'})^2 y_1^3 \frac{\beta k (3 \gamma^2 - 4 k^2)}{N} - \\[2.5ex]
		&(y_1^{'})^2 y_1^2 \gamma^2 k^2 ( \gamma + k) + 12 y_1^{'} y_1^5 \frac{\beta^2 k^2}{N^2} + \\[2.5ex]
		&y_1^{'} y_1^4 \frac{\beta \gamma k^2 (3 \gamma + 2 k)}{N} + 4 y_1^6 \frac{\beta^2 k^3}{N^2} + y_1^5 \frac{\beta \gamma^2 k^3}{N}=0
		&
	\end{aligned}
\end{equation}
 \underline{Input-output equation of the model \ref{Model3}}
\begin{equation} \notag 
	\begin{aligned}
    &(y_1^{'''})^2 y_1^{'} y_1 + (y_1^{'''})^2 y_1^2 k  - y_1^{'''}     (y_1^{''})^2 y_1 - y_1^{'''} y_1^{''} (y_1^{'})^2  + y_1^{'''} y_1^{''} y_1^{'} y_1 (2 \gamma - k) + \\[2.5ex]
    &2 y_1^{'''} y_1^{''} y_1^2 k (\gamma + k) - y_1^{'''} (y_1^{'})^3  (\gamma + k) + 4 y_1^{'''} (y_1^{'})^2 y_1^2 \frac{( - \beta_A \rho + \beta_A + \beta_I \rho)} {N \rho} +  \\[2.5ex]
    &y_1^{'''}  (y_1^{'})^2 y_1 (\gamma^2 - 2 k^2) +8 y_1^{'''} y_1^{'} y_1^3 \frac{k ( - \beta_A \rho + \beta_A + \beta_I \rho)}{N \rho}+ y_1^{'''} y_1^{'} y_1^2 \gamma k (2 \gamma + k) +  \\[2.5ex]
    &4 y_1^{'''} y_1^4 \frac{k^2 ( - \beta_A \rho + \beta_A + \beta_I \rho)}{N \rho} +y_1^{'''} y_1^3 \gamma^2 k^2 +(y_1^{''})^3 y_1^{'} - (y_1^{''})^3 y_1  (\gamma + k) + 2 (y_1^{''})^2 (y_1^{'})^2 k +   \\[2.5ex]
    &3 (y_1^{''})^2 y_1^{'} y_1^2 \frac{(\beta_A \rho - \beta_A - \beta_I \rho)}{N \rho} +(y_1^{''})^2 y_1^{'} y_1  (\gamma^2 - \gamma k - 2 k^2) +3 (y_1^{''})^2 y_1^3 \frac{k (\beta_A \rho - \beta_A - \beta_I \rho)}{ N \rho} +  \\[2.5ex]
    &(y_1^{''})^2 y_1^2 k  (\gamma^2 + 2 \gamma k + k^2) + y_1^{''} (y_1^{'})^3( - 2 \gamma^2 - \gamma k + 2 k^2)+ \\[2.5ex]
    &2 y_1^{''} (y_1^{'})^2 y_1^2 \frac{( - 3 \beta_A \gamma \rho + 3 \beta_A \gamma + \beta_A k \rho - \beta_A k + 3 \beta_I \gamma \rho - \beta_I k \rho)}{N \rho}+ \\[2.5ex]
    &y_1^{''} (y_1^{'})^2 y_1 (\gamma^3 - \gamma^2 k - 3 \gamma k^2 - 2 k^3)+2 y_1^{''} y_1^{'} y_1^3 \frac {k ( - 6 \beta_A \gamma \rho + 6 \beta_A \gamma - \beta_A k \rho + \beta_A k + 6 \beta_I \gamma \rho + \beta_I k \rho)}{N \rho} + \\[2.5ex]
    &y_1^{''} y_1^{'} y_1^2 \gamma k (2 \gamma^2 + 2 \gamma k + k^2) + 2 y_1^{''} y_1^4 \frac{k^2 ( - 3 \beta_A \gamma \rho + 3 \beta_A \gamma - 2 \beta_A k \rho + 2 \beta_A k + 3 \beta_I \gamma \rho + 2 \beta_I k \rho)}{N \rho} +  \\[2.5ex]
    &y_1^{''} y_1^3 \gamma^2 k^2(\gamma + k) + (y_1^{'})^5 \frac{ (\beta_A \rho - \beta_A - \beta_I \rho)}{N \rho} +  \\[2.5ex]
     &(y_1^{'})^4 y_1 \frac{ (4 \beta_A \gamma \rho - 4 \beta_A \gamma + 3 \beta_A k \rho - 3 \beta_A k - 4 \beta_I \gamma \rho - 3 \beta_I k \rho)}{N \rho} + \\[2.5ex]
    &( y_1^{'})^4 ( - \gamma^3 - 2 \gamma^2 k + k^3) + 4 (y_1^{'})^3 y_1^3 \frac{(\beta_A^2 \rho^2 - 2 \beta_A^2 \rho + \beta_A^2 - 2 \beta_A \beta_I \rho^2 + 2 \beta_A \beta_I \rho + \beta_I^2 \rho^2)}{N^2 \rho^2} + \\[2.5ex]
    &(y_1^{'})^3 y_1^2 \frac{ ( - \beta_A \gamma^2 \rho + \beta_A \gamma^2 + 6 \beta_A \gamma k \rho - 6 \beta_A \gamma k + 6 \beta_A k^2 \rho - 6 \beta_A k^2 + \beta_I \gamma^2 \rho - 6 \beta_I \gamma k \rho - 6 \beta_I k^2 \rho)}{N \rho} +\\[2.5ex]
    &(y_1^{'})^3 y_1 \gamma k  ( - 2 \gamma^2 - 3 \gamma k - k^2) + 12 (y_1^{'})^2 y_1^4 \frac{k (\beta_A^2 \rho^2 - 2 \beta_A^2 \rho + \beta_A^2 - 2 \beta_A \beta_I \rho^2 + 2 \beta_A \beta_I \rho + \beta_I^2 \rho^2)}{N^2 \rho^2} + \\[2.5ex]
 	\end{aligned}
\end{equation}
\begin{equation}\label{Model3_input_output}
	\begin{aligned}
	 &(y_1^{'})^2 y_1^3 \frac{k  ( - 3 \beta_A \gamma^2 \rho + 3 \beta_A \gamma^2 + 4 \beta_A k^2 \rho - 4 \beta_A k^2 + 3 \beta_I \gamma^2 \rho - 4 \beta_I k^2 \rho)}{N \rho} - (y_1^{'})^2 y_1^2 \gamma^2 k^2 (\gamma + k) + \\[2.5ex]
    &12 y_1^{'} y_1^5 \frac{k^2 (\beta_A^2 \rho^2 - 2 \beta_A^2 \rho + \beta_A^2 - 2 \beta_A \beta_I \rho^2 + 2 \beta_A \beta_I \rho + \beta_I^2 \rho^2)}{N^2 \rho^2} + \\[2.5ex]
    &y_1^{'} y_1^4 \frac{ \gamma k^2 ( - 3 \beta_A \gamma \rho + 3 \beta_A \gamma - 2 \beta_A k \rho + 2 \beta_A k + 3 \beta_I \gamma \rho + 2 \beta_I k \rho)}{N \rho} + \\[2.5ex]
    &4 y_1^6 \frac{ k^3 (\beta_A^2 \rho^2 - 2 \beta_A^2 \rho + \beta_A^2 - 2 \beta_A \beta_I \rho^2 + 2 \beta_A \beta_I \rho + \beta_I^2 \rho^2)}{N^2 \rho^2} + y_1^5 \frac{ \gamma^2 k^3 ( - \beta_A \rho + \beta_A + \beta_I \rho)}{N \rho}
	\end{aligned}
\end{equation}

\bigskip
\bibliographystyle{apalike}
\bibliography{references_aug23}
  \end{document}